\documentstyle[preprint,aps]{revtex}
\begin{document}
\title{The $^7$Be Solar Neutrino Line:\\
A Reflection of the Central Temperature Distribution\\
 of the Sun.}

\author{John N. Bahcall}

\address{Institute for Advanced Study, Princeton, NJ 08540}

%\date{}

\maketitle

\begin{abstract}
A  precise test of the theory of
stellar evolution can be performed
by measuring the average difference in energy between the
neutrino line produced by ${\rm ^7Be}$ electron capture in the solar
interior and the corresponding
neutrino line produced in a terrestrial laboratory.
This energy shift is calculated to be 1.29~keV (to an accuracy of a few
percent) for the dominant ground-state to ground-state transition.
The energy shift is approximately equal to the average temperature of the
solar core, computed by integrating the temperature over the solar interior
with a weighting factor equal to the
locally-produced $^7$Be neutrino emission.
Therefore, a measurement of the energy shift is
a measurement of the central temperature distribution of the sun.

The energy profile of the ${\rm ^7Be}$ line is derived analytically
and is evaluated numerically.  The line shape is asymmetric: on the
low-energy side,
the line shape is Gaussian with a half-width
at half-maximum of 0.6 keV and on the high-energy side, the line shape is
exponential
with a half-width at half-maximum
of 1.1 keV.  The effective temperature of the
high-energy exponential tail is $15 \times 10^6$ K.
The energy profile of the $^7$Be neutrino line should be taken into
account in calculations of vacuum neutrino oscillations
and of the absorption cross section for $^7$Be
solar neutrinos incident on $^7$Li nuclei.

The characteristic modulation
of the ${\rm ^7Be}$ line shape that would be caused by either vacuum neutrino
oscillations or by
matter-enhanced (MSW) neutrino oscillations is shown to be small.
Other frequently-discussed weak interaction solutions to the solar
neutrino problem are also not expected to change significantly the line
profile.
\end{abstract}


\section{Introduction}
\label{sec:introduction}

\subsection{Background}
\label{sec:background}

The original motivation\cite{Bahcall64}
for performing solar neutrino experiments
was to learn more about how the sun shines.
When the first observational results from the chlorine experiment became
available\cite{Davis68},
the focus shifted\cite{BahcallDavis76}
from learning about the interior of the sun
to trying to determine if the discrepancy
between calculation and observation
was due to inadequacies in the astrophysics
or to new weak interaction physics.
Directed towards this goal, four experiments are being performed
\cite{Davis89,Hirata91,Anselmann92,Abazov91} to measure the
fluxes of neutrinos that are produced by nuclear
fusion reactions in the solar interior.  An additional four
solar neutrino
experiments\cite{Aardsma87,Totsuka90,Raghavan90,Bahcall86}---designed
to determine if new physics or new astronomy is required---are being developed.

Progress toward the goal of testing solar theory by measuring solar
neutrinos has been complicated by what may be the discovery
of new weak interaction physics.
A comparison of two of the existing experiments, the
chlorine\cite{Davis89} and the
Kamiokande\cite{Hirata91} experiments, suggests
\cite{BahcallBethe} the existence of
a physical process that changes---in a way that depends upon neutrino
energy---the fraction of electron-type neutrinos that reach detectors on earth
after being created in the core of the sun.

At first glance,
the terrestrially-observed flux of $^8$B neutrinos is encouragingly
close to the flux calculated on the basis of the standard solar model
and the standard electroweak theory, especially considering the
sensitivity of the predicted flux to details of the stellar physics.
However, the discrepancy between calculation and observation is significant
since the theoretical uncertainties are smaller than the difference
between what is measured and what is
predicted\cite{Bahcall89,Bahcall88,Bahcall92}.
This discrepancy has stimulated a number of imaginative and
attractive possible
interpretations in terms of new physical
processes
\cite{Gribov69,Wolfenstein78,Mikheyev86,Roulet91,Cisneros71,Lim88,Pal92}.
Until the
effects of these proposed new physical processes are either established or
rejected experimentally, quantitative
astronomical inferences from the measured magnitudes of the
solar neutrino fluxes will be
limited.

\subsection{The $^7$Be Energy Profile}
\label{sec:beenergyprofile}

The purpose of this article is to draw attention to precise
new predictions of the theory of stellar evolution that can be tested
with the aid of future solar neutrino experiments. These new predictions
are based upon the calculated shape of the energy
spectrum of neutrinos produced by ${\rm ^7Be}$ electron
capture in the solar interior.
The reaction in question is
\begin{equation}
e^- + ^7{\rm Be}\  \to\ ^7{\rm Li} + \nu_e\ .
\label{capturereaction}
\end{equation}
In the interior of the sun, most of the electrons are captured from
continuum (unbound)states\cite{Bahcall62,Iben67}.  The electron-capture
reaction shown in (\ref{capturereaction}) produces a neutrino line
because the $^7$Li nucleus in the final state has a mass that is much
greater than the energy of the neutrino.  The recoiling nucleus takes up
a significant amount of momentum, but only a negligible amount of energy.
The focus of this paper is on the broadening by thermal effects of
the neutrino line produced in the sun.

This paper shows that the line shape for Eq.\ (\ref{capturereaction})
reflects accurately the temperature distribution in the interior of
the sun.
As we shall see,
the difference between the average energy of the neutrino line emitted
in the sun and in the laboratory  is approximately equal to the central
temperature of the sun (see especially
Sec.~\ref{sec:energyspectrum} and Sec.~\ref{sec:energyshift} ).
The predicted 1.3 keV
increase in the average energy of the solar ${\rm ^7Be}$ neutrino line
relative to laboratory decays is the simplest quantity to measure that
directly reflects the solar temperature distribution.

In addition to a shift in the average neutrino
energy, the shape of the line profile for
${\rm ^7Be}$ electron capture in the sun is
different from the line profile that would be observed for a laboratory
source of ${\rm ^7Be}$ neutrinos.
The shift in average energy and the change in the shape of the line profile
are both caused by the high temperatures in
the core of the sun where the neutrinos are produced.
The high solar temperatures produce
significant thermal energies for continuum electrons and their
capturing nuclei, Doppler shifts of the emitting nuclei,
a high degree of ionization of solar ${\rm ^7Be}$ ions, and
a difference in atomic binding energies for solar
and laboratory ${\rm ^7Be}$ atoms.\footnote{After the initial version of
the present work was accepted for publication\cite{Bahcall93}, my
attention was directed by J.
Pantaleone to two important papers\cite{Pakvasa90}. In these
two papers, the effect of the line
width of the $^7$Be and the $pep$ lines on neutrino oscillations was
calculated.}

Figure 1 shows the two nuclear transitions that occur when ${\rm ^7Be}$
captures an electron in the laboratory.  The neutrino energy
corresponding to the transition from the ground-state of ${\rm ^7Be}$ to the
ground-state of ${\rm ^7Li}$ will be denoted by
$q_{\rm Lab}({\rm g.s.})$;
the energy corresponding to the transition
to the first excited state of ${\rm ^7Li}$ will be denoted by
$ q_{\rm Lab} ({\rm ex. s.}) $.  In what follows, I will discuss
both transitions on an equal basis since the physical processes
determining the neutrino line shape are the same in both
cases.  However, the transition to the ground state of ${\rm ^7Li}$ is
more easily studied experimentally because it has
a higher energy and a larger branching ratio\cite{Lederer78}.
Both the higher energy
and the larger branching ratio
of the ground-state to ground-state transition
contribute to making the
expected rate for this transition faster by an order of magnitude
than for the ground-state to
excited-state transition.

For convenience, I will refer
to the ground-state to ground-state
transition as ``ground-state'' capture and will refer to the
ground-state to excited-state transition as ``excited-state'' capture.
In an additional effort to avoid cumbersome phrasing, I will often refer in
the singular to ``the'' line profile or to ``the'' energy shift, when I
mean the line profile or the energy shift for both ground-state
capture and excited-state capture.

Before proceeding to the calculations, it is helpful to think
about the following question.  Why is
the effect of the solar environment on the shape of the energy spectra
for continuum beta-decays (one particle in
the initial state, three particles in the final state, e.g., $^8$B decay)
different
from the effect of solar conditions on the profile shape for
the two-body electron capture reactions?  This question has a simple
physical answer.

If the simplest version of the standard electroweak model is correct,
then the shape of the energy spectrum for electron-type neutrinos
from continuum
beta-decays in the sun, such as the
$^8$B beta-decay, is independent of the conditions in the sun to an
accuracy of 1 part in $10^5$\cite{Bahcall91}.
The invariance is a result of the fact that,
in the center-of-momentum frame, a decaying $^8$B nucleus has no
kinetic energy.
In the laboratory frame, terms of order the velocity of the nucleus,
$v({\rm ^8B})/{\rm c}$, cancel out because there are as many $^8$B
nuclei moving toward the observer as there are moving away from the
observer.
The first-order Doppler effects vanish because at each point
within the broad $^8$B continuum the neutrino energy is spread out
symmetrically by a small amount.
The largest potentially-observable effects
of the solar temperature
on the observed $^8$B neutrino energy spectrum are only second
order terms,$\sim v({\rm ^8B})^2/{\rm c}^2$\cite{Bahcall91}.
This implication of standard theory will be tested by
experiments\cite{Aardsma87,Totsuka90,Bahcall86}
that will measure the
shape of the $^8$B neutrino energy spectrum and will compare the
observed shape with the spectrum determined from terrestrial measurements.

By contrast,
in the two-body electron-capture reactions,
the profile of a narrow neutrino emission
line is broadened asymmetrically by the solar temperature.  On the
low-energy side of the
neutrino line profile,
the dominant effect  is
first-order Doppler (or Gaussian) broadening caused by the motion of the
decaying nuclei.
On the high-energy side of the observed profile, the dominant effect is
an exponential broadening resulting from the center-of-momentum kinetic energy
of the electron and the decaying nucleus
(for a physical discussion of the profile shape see especially
Sec.\ \ref{sec:approximate}).
This asymmetric broadening
causes a shift of the average energy of the neutrino line as well as
change in the shape of the line profile.

\subsection{Experimental Possibilities}
\label{sec:hope}

A number of experiments\cite{Raghavan90,Drukier84} have been proposed that
would measure predominantly the $\nu_e$ flux from $^7$Be electron
capture in the sun, using
detectors that are based upon neutrino-electron scattering.
At the present time,
the BOREXINO experiment\cite{Raghavan90}
is the most advanced of these proposals and can, if recent estimates of the
expected backgrounds are correct, measure
the flux of
$^7$Be neutrinos.  Radiochemical detectors of the flux of $^7$Be neutrinos
have also been discussed\cite{Rowley78}, but these
detectors do not give specific information on neutrino energies.
The radiochemical detectors are efficient for measuring the total flux of
electron-type
neutrinos above a specified energy threshold, but are not useful for
studying the thermal effects investigated in this paper. Most recently,
the use of a high resolution LiF bolometric detector of $^7$Be neutrinos
has been discussed \cite{Raghavan93}.

Detectors have been developed\cite{Caldwell88} for a variety of
other applications, including dark matter searches, the observation of double
beta decay, and x-ray astronomy, that have the energy resolution and
the sensitivity that are
required to study the energy spectrum of the $^7$Be neutrino
line.
The best-available detectors have
energy resolutions, ${\Delta E/E}$,
of much better than $1\%$, but they are smaller than
would be required for a full-scale solar neutrino detector.

The most direct way to study the $^7$Be energy profile may be to
detect neutrino absorption by nuclei, a process which leaves an electron and a
recoiling nucleus in the final state.
In these charged-current
transitions, nearly all of the
initial neutrino energy is transferred to the final-state electron
(the nuclear recoil energy being small).
In the neutrino-electron scattering experiments that are currently under
development, the 1 keV width of the $^7$Be line is spread out over
several hundred keV of electron recoil energy, since the neutrino and
the electron share the final state state energy  (see Figure 8.5 of
\cite{Bahcall89}).

In order to measure the predicted 1.3 keV ($0.15$\%) energy shift via neutrino
absorption, an energy
resolution of order $1\%$ to $0.1\%$ is desirable,
depending somewhat upon the absorption threshold.
Consider,
for specificity, a conceivable cryogenic experiment\cite{Alessandrello92}
that might be performed on $^{81}$Br with an energy resolution of $1\%$ and
with a total of
$10^3$ measured neutrino events. The energy released to the recoil
electron would be about $400$ keV (the reaction threshold is about $450$
keV), so the average neutrino energy would
be measured to an accuracy of about 0.1 keV.
The experimental parameters assumed in the above discussion
would permit one
to measure the central temperature of the sun to an accuracy of about 10\%.
In addition, a proposed high-pressure helium gas
detector\cite{Sequinot92} might well have sufficient energy resolution
to measure the predicted energy shift.

The requirements for a practical experiment may
be achievable since the solar neutrino detectors currently under
development are designed to detect several thousand events per year (albeit
with much poorer energy resolution, typically $\sim 10\%$).
It would be valuable to calibrate the solar results by studying an
intense laboratory source of $^7$Be neutrinos with the same detector as
used in the solar observations.
The work described in this paper was undertaken in the hope that it
would stimulate an experiment that would measure the energy
shift and perhaps other characteristics of the $^7$Be line profile,
in somewhat the same way that
the initial theory and the experimental results on solar neutrinos developed
together\cite{Bahcall64}.

 \subsection{Organization and Previous Work}
\label{sec:organization}

This paper is organized as described below.  Section~\ref{sec:energies}
 presents
calculations of the average neutrino energy release
in the rest frame of the decaying
particle when a $^7$Be nucleus captures an electron under laboratory
conditions (see Sec.~\ref{sec:labdecays}) and in a solar environment (in
which most of the electrons are captured from continuum orbits, see
Sec.~\ref{sec:solardecays}).  Section~\ref{sec:continuumcapture}
describes the central calculation of this paper, an evaluation of the
energy profile for the neutrino line emitted when $^7$Be nuclei capture
continuum electrons that have a specified temperature.  The following
section, Sec.~\ref{sec:boundcapture}, outlines the calculation of the
energy profile for the small but significant fraction of the solar
electron captures that occur from bound orbits.  The numerical
characteristics of the line shape, including the shift in average
neutrino energy
between laboratory and solar decays, the full-width at half-maximum of
the line profile, and
the lower-order moments of the energy distribution, are the subject of
Sec.~\ref{sec:energyprofile}.  The line profile is averaged, in this
section, over the physical characteristics, including the temperature
distribution, of detailed solar models.  The numerical results are
presented in Tables I and II and displayed in Figures 2 and 3.
Section~\ref{sec:approximate} provides approximate
analytic derivations of the low-energy  half-width (0.56~keV) and the
high-energy  half-width (1.07~keV) of the neutrino energy profile and
isolates the separate physical origins of these two features.  This
section also presents a derivation of the shift in average neutrino
energy between
solar and laboratory decays that explains why the shift is the same for
ground-state and for excited-state captures.
Section~\ref{sec:othereffects} evaluates the effects on the neutrino
line profile of electrostatic screening,
of gravitational redshifts, and of collisional broadening, and shows that
these effects are small.  The influence of possible new weak interaction
physics, exemplified by vacuum oscillations and by the MSW effect, is
the subject of Sec.~\ref{sec:newphysics}.  Neither type of oscillation
would affect significantly the measurable characteristics of the
$^7$Be line profile.
The energy shift and the asymmetric profile of the $^7$Be line do
change the
computed absorption cross section for $^7$Be solar neutrinos incident on
a $^7$Li detector, as is shown in Section~\ref{sec:Li7neutrino}.
Section~\ref{sec:summary} summarizes
the principal results of this paper.

There have been a number of previous calculations of the total rate for
$^7$Be electron capture in the sun over the past three decades
\cite{Bahcall62,Iben67,Bahcall69,Watson73,Domogatsky69,Bahcall78,Johnson92}.
The present work on the $^7$Be line shape does not change the results
of these prior calculations of the total capture rate, the standard
value in general use still being the one given in reference\cite{Bahcall69}.
The only earlier work on the
 broadening of the $^7$Be neutrino line with which I am familiar
concentrated either on the fraction of electron-capture
neutrinos that were above threshold for the $\nu_e + {\rm ^7Li} \to {\rm
^7Be} + e$ reaction, taking account of center-of-momentum energies but
not Doppler shifts \cite{Domogatsky69,Bahcall78}, or on a crude estimate of
the total width, based only on the Doppler shifts of the $^7$Be ions
\cite{Bahcall89}.

\section{Laboratory and Solar Energies}
\label{sec:energies}

The average neutrino energy that is
released when ${\rm ^7Be}$ captures an electron in
a terrestrial
laboratory is different from the average energy that is released when
${\rm ^7Be}$ captures an electron in the core of the sun.  In a laboratory
experiment, all of the captured electrons are initially bound in a
${\rm ^7Be}$ atom, whereas in the solar interior most of the electron captures
occur from continuum orbits\cite{Bahcall62}.
Part of the difference in the average neutrino
energy release, the part on which this section concentrates, is due,
therefore,
to the different atomic binding energies
in the laboratory and in the sun.

Energy conservation implies, for electron captures either in the sun or in the
laboratory, that

\begin{equation}
E_{\rm initial} - E_{\rm final}  = \Delta M + K(e) + K({\rm ^7Be})  -
K({\rm ^7Li})- q + a(^7{\rm Be}) - a(^7{\rm Li}) = 0 ,
\label{energyconservation}
\end{equation}
where $E_{\rm initial}$ and $E_{\rm final}$ are
the total energies in the initial and final states,
$K(e)$ and $K({\rm ^7Be})$ are kinetic energies of the electron and of
the ${\rm ^7Be}$ nucleus in the initial state,
$K({\rm ^7Li})$ is the kinetic energy
of the recoiling  ${\rm ^7Li}$ nucleus, $q$ is the neutrino energy, and
$a(^7{\rm Be})$ and $a(^7{\rm Li})$ are the atomic binding energies.
The quantity $\Delta M$ in Eq.\ (\ref{energyconservation}) is
the difference in rest mass energies \cite{Lederer78}, excluding atomic
binding energies,

\begin{equation}
\Delta M = m_e + M({\rm ^7Be}) - M({\rm ^7Li}) = 862.10~{\rm keV} .
\label{massdelta}
\end{equation}
For electron captures that occur in the laboratory, the initial kinetic
energies, $K(e)$ and $K({\rm ^7Be})$, are both zero.  However, these terms
contribute numerically the largest amount to the difference between
laboratory and solar neutrino release and will be
calculated in Secs.\ \ref{sec:continuumcapture} and \ref{sec:boundcapture}.

In the following two subsections, I evaluate the neutrino energy using
Eq.\ (\ref{energyconservation}) for laboratory decays (in Sec.\
\ref{sec:labdecays})
and for solar decays (in Sec.\ \ref{sec:solardecays}).
I discuss in Sec.\ \ref{sec:electroenergy}
the electrostatic energy of the screening cloud that surrounds
the decaying nucleus.

\subsection{Laboratory Decays}
\label{sec:labdecays}

The laboratory neutrino energy, $q_{\rm Lab}$, satisfies
the equation

\begin{equation}
q_{\rm Lab} + q^2_{\rm Lab}/2M ({\rm ^7Li})= \Delta M +
a_{\rm Lab} ({\rm ^7Be}) - a_{\rm Lab} ({\rm ^7Li}),
\label{qlabgeneral}
\end{equation}
where the difference in atomic binding energies is

\begin{equation}
a_{\rm Lab} ({\rm ^7Be}) - a_{\rm Lab} ({\rm ^7Li}) = - 0.195~{\rm keV}.
\label{atomiclab}
\end{equation}
Since the energy difference between an initial and a final state is
independent of the choice of steps used to reach the final state,
the difference in atomic binding energies is equal to the
difference between the sum of the successive ionization potentials
of $^7$Be and the sum of the successive ionization potentials of $^7$Li.
The difference in atomic binding energies given in Eq.\ (\ref{atomiclab})
was calculated by subtracting the sum of the three ionization potentials
of the lithium atom from the sum of
the four ionization potentials of atomic beryllium, using the
measured values for the ionization potentials\cite{Handbook79}.

Combining Eq.\ (\ref{massdelta}) and Eq.\ (\ref{atomiclab}), one obtains
\begin{equation}
m_e + M ({\rm ^7Be}) - M ({\rm ^7Li}) + a_{\rm Lab} ({\rm ^7Be}) - a_{\rm Lab}
({\rm ^7Li}) = 861.90~{\rm keV} ,
\label{sumlab}
\end{equation}
which is the tabulated mass difference of the neutral atoms
\cite{Lederer78}.
The small contribution of the ${\rm ^7Li}$ recoil energy is
\begin{equation}
q^2 ({\rm g.s.})/2M({\rm ^7Li}) \cong 0.057~{\rm keV}\ ;\ q^2({\rm ex.s.})/2M
(^7{\rm Li}) = 0.011~{\rm keV},
\label{recoillab}
\end{equation}
for ground-state and excited state transitions, respectively.
Inserting Eq.\ (\ref{sumlab}) and Eq.\ (\ref{recoillab})
in Eq.\ (\ref{qlabgeneral}) , we find the
neutrino energies for the laboratory transitions,
\begin{equation}
q_{\rm Lab}({\rm g.s.}) \cong 861.84~{\rm keV},
\label{gsenergylab}
\end{equation}
and
\begin{equation}
q_{\rm Lab} ({\rm ex.
s.}) = 384.28~{\rm keV},
\label{esenergylab}
\end{equation}
that are shown in Fig. 1.

\subsection{Solar Decays}
\label{sec:solardecays}

This subsection begins with a calculation of the neutrino
energy that is released in the limiting case
in which a continuum electron with zero kinetic
energy is captured by a ${\rm ^7Be}$ nucleus at rest.  I then calculate the
neutrino energy that is emitted when a bound electron is captured from a
stationary ${\rm ^7Be}$ nucleus.
I evaluate
the influence of non-zero kinetic energies of the electron and of the
${\rm ^7Be}$ ion in Secs.\ \ref{sec:continuumcapture} and
\ref{sec:boundcapture}.

For the special case of continuum electron capture at rest,
the neutrino energy release is
the same as for laboratory capture except that the nucleus is assumed
stripped of all bound electrons.  Referring to Eq.\ (\ref{energyconservation}),
one can write:
\begin{equation}
q_{\rm cont,star} = q_{\rm Lab} -
[ a_{\rm lab} ({\rm ^7Be}) - a_{\rm lab} ({\rm ^7Li})].
\label{energycontstar}
\end{equation}
Therefore,
\begin{equation}
q_{\rm cont,star} = \Delta M - q^2_{\rm cont,star}/2M ({\rm ^7Li}),
\label{approximatecontstar}
\end{equation}
where the difference, $\Delta M$, in rest mass energies is given by
Eq.\ (\ref{massdelta}).
Thus
\begin{equation}
q_{\rm cont,star} ({\rm g.s.}) = 862.04~{\rm keV}\ .
\label{contstarvalue}
\end{equation}

I next consider the average
neutrino energy that is released, $q_{\rm bound,star}$,
when an electron is captured from a bound orbit. The only difference
between $q_{\rm bound,star}$ and the previously calculated
$q_{\rm cont,star}$ is the atomic binding energy of the electrons.
Therefore, one can write
\begin{equation}
q_{\rm bound,star} = q_{\rm cont,star} +\langle a_{\rm star} (^7{\rm Be}) -
a_{\rm star}(^7{\rm Li}) \rangle ,
\label{boundcont}
\end{equation}
where the angular brackets denote an average over the sun.
The K-shell binding energy has been determined for solar conditions
by Iben and his
collaborators\cite{Iben67}using a variational calculation.
For the case in which one
electron is bound to the ${\rm ^7Be}$ nucleus, the result can be expressed in
terms of the relative K-shell
binding, $\sigma_R$, in the sun compared to the laboratory value, where
$\sigma_R$ is defined by the following equation:
\begin{equation}
a_{\rm K,star} (^7{\rm Be}) = - 216.6 \sigma_R~{\rm eV} .
\label{sigmaRdefinition}
\end{equation}

At the peak of the ${\rm ^7Be}$ solar neutrino emission,
$\sigma_R \cong 0.25$ and
$a_{\rm K,star} (^7{\rm Be}) = -0.06$~keV (cf. discussion in
Sec.~\ref{sec:energyprofile}).  This binding energy is
sufficiently small that it is not necessary, for our purposes, to
calculate $a_{\rm K,star} (^7{\rm Be})$ to high precision.  However, an
accurate calculation was carried out without difficulty and the results
of this calculation are described below. The average binding energies
were obtained for two standard solar models\cite{Bahcall92} (see also
Sec.~\ref{sec:solarmodels}), one that
included, and one that did not include, helium diffusion.
Using the tabulated results of the variational-principle calculation
\cite{Iben67},
a convenient interpolation formula was derived\cite{Bahcall69} for $\sigma_R$
in terms of the local density, temperature, and chemical
composition.
Weighting the value of $\sigma_R $ calculated at a given radial distance
from the center of the sun by the
${\rm ^7Be}$ neutrino flux at that radius, the average relative
K-shell binding energy for the
standard solar model\cite{Bahcall92}  including helium diffusion is
\begin{equation}
\sigma_R \simeq 0.255,
\label{sigmaRvalue}
\end{equation}
and is 0.256 for the solar model that does not include helium diffusion.
Thus, the average binding energy of a K-shell electron in the sun is
\begin{equation}
\langle a_{\rm K,star}(^7{\rm Be})\rangle \cong -0.055~{\rm keV} .
\label{asubK}
\end{equation}
The difference in atomic binding energies between two K-shell electrons
in a beryllium atom in the sun
and a single K-shell electron in a lithium
atom in the sun may be estimated to more-than-adequate accuracy by
scaling the K-shell energy given in Eq.\ (\ref{asubK}) by the
ratio of the square of the lithium and the beryllium nuclear charges.
One can also estimate the energy difference by scaling
(see \cite{Iben67}), using the same value of $\sigma_R$
calculated above,
the laboratory energy that is required to remove one
electron from previously twice-ionized beryllium, leaving behind one electron
bound to a lithium atom.
Both methods give the same answer
to the accuracy of interest here.  One finds
\begin{equation}
\langle a_{\rm 2K,star}(^7{\rm Be})- a_{\rm K,star}(^7{\rm Li})\rangle \
\simeq\  -0.079~{\rm keV} .
\label{asub2K}
\end{equation}
One must average the results given in Eq.\ (\ref{asubK}) and
Eq.\ (\ref{asub2K}) over the stellar model taking account of the
variable fraction of the decays
that occur from ${\rm ^7Be}$ atoms with one or with
two bound electrons.
Let $p_1$ and $p_2$ be the respective probabilities that $^7{\rm Be}$
has one or two bound electrons that can decay by K-capture.
Then the appropriate average can be written as

\begin{equation}
\langle a\left(^7{\rm Be}\right) - a\left(^7{\rm Li}\right)\rangle_{\rm
star} =
\langle \frac{p_1 a_{\rm K,star}\left(^7{\rm Be}\right) + 2p_2\left(a_{\rm
2K,star}\left(^7{\rm Be}\right) - a_{\rm 2K,star}\left(^7{\rm
Li}\right)\right)}{p_1 + 2p_2} \rangle
\label{}
\end{equation}
Convenient expressions for $p_1$ and $p_2$ have been given by Iben and
his collaborators\cite{Iben67}.
Carrying out the average using the results for the relative ionization
states obtained from the variational-principle calculation, I find
for the atomic binding energy in the sun:
\begin{equation}
\langle a(^7{\rm Be}) - a(^7{\rm Li})\rangle_{\rm star}\  =\ -0.06~{\rm
keV}.
\label{solaratomicbinding}
\end{equation}
Inserting Eq.\ (\ref{solaratomicbinding}) in Eq.\ (\ref{boundcont}),
\begin{equation}
q_{\rm bound,star}({\rm g.s.})  = 861.98~{\rm keV}.
\label{qboundstargs}
\end{equation}

By an analogous procedure, one can obtain the following results for
the excited-state transition:
\begin{equation}
q_{\rm cont,star}({\rm ex.s.}) = 384.43~{\rm keV},
\label{qcontes}
\end{equation}
and
\begin{equation}
\ q_{\rm bound,star} ({\rm ex.s.}) =
384.38~{\rm keV}.
\label{qboundstares}
\end{equation}

\section{Capture from Continuum Orbits}
\label{sec:continuumcapture}

This section presents a calculation of the
energy profile for the neutrino line emitted when
 electrons are captured from continuum orbits by ${\rm ^7Be}$ ions
in the core of the sun.  The energy profile is the
probability that  a ${\rm ^7Be}$ ion captures an electron from the continuum
and emits a neutrino of any specified  energy.
As noted earlier, continuum electron capture is the
dominant process by which ${\rm ^7Be}$ nuclei decay in the solar interior
\cite{Bahcall62}.
The following section, Sec.\ \ref{sec:boundcapture},
presents a calculation of the line profile when
bound electrons are captured by ${\rm ^7Be}$ ions in the solar interior.

The partial transition probability to undergo electron capture can be
written schematically,
for a given relative
flux density of electrons
and ${\rm ^7Be}$ ions, as
\begin{equation}
d\lambda = {\rm flux} \times \sigma_{\rm cap},
\label{schematic}
\end{equation}
where $\sigma_{\rm cap}$ is the appropriate weak interaction
capture cross section. More
specifically,  the partial transition probability can be written in
terms of the usual\cite{Konopinski66}
charged-current beta-decay Hamiltonian, $H_{\beta}$,
as
\begin{equation}
d\lambda =  2\pi \vert
\langle f\vert H_\beta \vert i\rangle\vert^2 \delta\left(E_{\rm initial}
 - E_{\rm final}\right),
\label{capHbeta}
\end{equation}
where the continuum wave functions are assumed to be normalized in a
large, finite volume.
Averaging over initial states, summing over final states, and
integrating over all space, one obtains
\begin{equation}
{\rm flux}\ \times\ d ({\rm cross\ section})
= (2\pi)^{-2}\sum_f
\overline{\sum_i} \vert \langle f\vert H_\beta\vert i\rangle\vert^2\delta^{(4)}
(p_i - p_f) d^3 \roarrow{q} d^3 \roarrow{p_7},
\label{vsigma}
\end{equation}
where the momenta in the final state of the neutrino and of the ${\rm ^7Li}$
ion
are denoted by $\roarrow{q}$ and by $\roarrow{p_7}$, respectively.
The beta-decay Hamiltonian can be written as
\begin{equation}
H_\beta = 2^{-1/2}G \left(\bar\psi_\nu \gamma_\alpha \left(1 +
\gamma_5\right)\psi_e\right)\left(\bar\psi_n \gamma_\alpha\left(C_V -
C_A \gamma_5\right)\psi_p\right) ,
\label{usualbeta}
\end{equation}
where all the symbols have their usual meanings
\cite{Konopinski66,Konopinski59,Bahcallapj64}.

It is convenient to carry out the calculations in the center-of-momentum
coordinate system.  Relative to the laboratory frame in which the
electron has a momentum $\roarrow{p_e}$ and the ${\rm ^7Be}$ ion has a momentum
$\roarrow{p_7}$, the center-of-momentum frame has a velocity
$\roarrow{V}$ given
by:
\begin{equation}
\roarrow{V} = M^{-1}  \left( \roarrow{p_e} + \roarrow{p_7}\ \right),
\label{velocityCM}
\end{equation}
where
\begin{equation}
M = m_e + M ({\rm ^7Be}).
\label{summedmass}
\end{equation}
The equation describing the conservation of energy
can be expressed simply in terms of
the momentum, $\roarrow{p}$, in the center-of-momentum frame,
where $\roarrow{p}$ is given by
\begin{equation}
\roarrow{p} = M^{-1}\left(M({\rm ^7Be})\roarrow{p_e} - m_e
\roarrow{p_7}\right) .
\label{definitionp}
\end{equation}
One can rewrite Eq.\ (\ref{energyconservation}) in the form:
\begin{equation}
E_{\rm initial} - E_{\rm final} \  =\  \Delta M + \frac{p^2}{2\mu}- q -
\frac{q^2}{2 M (^7{\rm Li})} ~=~ 0,
\label{reenergyconservation}
\end{equation}
where the reduced mass $\mu$ is
\begin{equation}
\mu = \frac{m_e M({\rm ^7Be})}{m_e + M ({\rm ^7Be})} .
\label{reducedmass}
\end{equation}
The solution of Eq.\ (\ref{reenergyconservation}) yields an expression
for the neutrino energy in the center-of-momentum frame in terms of the
momentum, $p$, of either of the particles, i.e.,
\begin{equation}
q \equiv q(p).
\label{qofp}
\end{equation}

The beta-decay matrix element in Eq.\ (\ref{vsigma}) can be greatly
simplified by using the so-called ``normal
approximation''\cite{Konopinski66,Konopinski59,Bahcallapj64}
that gives the leading term for the matrix element
in a rapidly convergent power series expressed
in terms of the small $(\lesssim 0.01)$ quantities $q R_{\rm nucleus}$ and
$p R_{\rm nucleus}$, where $R_{\rm nucleus}$ is the nuclear radius
of ${\rm ^7Li}$.
The conditions for the applicability of the normal
approximation in stars are given in Eq. (10) of ref. \cite{Bahcallapj64}
and are satisfied to high accuracy in the present case.

Making the
usual non-relativistic reduction and summing over the spins of the
nuclei in the initial and final states and over the spins of the
electron and the
neutrino, one finds in the center-of-momentum frame
a simple expression for the product of the
relative velocity of the particles times the capture cross section,
namely,
\begin{equation}
v_{\rm rel}\sigma_{\rm cap} = \frac{G^2 \xi\vert \psi_e\vert^2 q^2(p)}{2\pi
\left[1 + q(p)/M(^7{\rm Li})\right]},
\label{reducedvsigma}
\end{equation}
where $\vert \psi_e\vert^2 $ is the electron probability density
(enhanced by the Coulomb attraction) averaged over the nucleus and
$\xi$ is the familiar sum of reduced matrix elements that occurs
in allowed beta-decays\cite{Konopinski66,Konopinski59,Bahcallapj64}, i. e.,
\begin{equation}
\xi = C^2_V \langle 1\rangle^2 + C^2_A \langle \sigma\rangle^2 ,
\label{definitionxi}
\end{equation}
where $\langle 1\rangle$ and $\langle\sigma\rangle$ refer, respectively,
to Fermi and Gamow-Teller matrix elements.

In a spherical shell at temperature $T$ that contains $N(^7{\rm Be})$
total ${\rm ^7Be}$ ions, the rate at which ${\rm ^7Be}$ captures electrons is
\begin{equation}
d\  {\rm Rate} (T) = N(^7{\rm Be}) \langle n (e)
\sigma_{\rm cap} v_{\rm rel}\rangle ,
\label{drate}
\end{equation}
where the average that is indicated in Eq.\ (\ref{drate}) is over the
thermal distributions of the electrons and the ${\rm ^7Be}$ ions.
Writing out the thermal average explicitly, one finds
\begin{eqnarray}
d\  {\rm Rate} (T) &&= \frac{d V n (^7{\rm Be}) n (e) G^2 \xi}{2 \pi
}\left(\frac{m_e}{2\pi kT}\right)^{3/2}\left(\frac{M
(^7{\rm Be})}{2 \pi kT}\right)^{3/2}\nonumber \\
&&\times \int\int d^3 p_e d^3 p_7 \exp\left(- p^2_e/2m_e
kT\right)\exp\left(- p^2_7/2M (^7{\rm Be}) kT\right)\nonumber \\
&&\times \frac{\vert \psi_e\vert^2 q^2(p)}{\left[1 + q/M(^7{\rm
Li})\right]} ,
\label{thermalaverage}
\end{eqnarray}
where $dV$ is the volume of the spherical shell and $n(e)$ and $n({\rm ^7Be})$
are, respectively,  the local number density of electrons and of ${\rm ^7Be}$
ions.
Converting the integration variables to the center-of-momentum
quantities, $\roarrow{p}$ and $\roarrow{P} = M \roarrow{V}$
[see Eq.\ (\ref{velocityCM})  and Eq.\ (\ref{definitionp})], one obtains
the relatively simple looking equation,
\begin{equation}
d\  {\rm Rate} (T) = C (T) \int\int d^3 \roarrow{p} d^3 \roarrow{P}
\exp \left(- p^2/2\mu
kT\right)\exp\left(- P^2/2M kT\right)
\frac{\vert \psi_e\vert^2 q^2(p)}{\left[1 + q(p)/M (^7{\rm Li})\right]},
\label{simplelooking}
\end{equation}
where $C(T)$ is a constant that is independent of neutrino energy.
To high accuracy in the present application,
one can approximate the electron probability density
by the non-relativistic expression\cite{Konopinski66,Bahcallapj64},
\begin{equation}
\vert \psi_e\vert^2 \cong \frac{8 \pi \alpha v_e^{-1}}{1 - \exp \left(-8\pi
\alpha v_e^{-1}\right)}\ ,\ \ \ v_e = \mu v_{\rm rel}/m_e  ,
\label{xinonrelativistic}
\end{equation}
where $\alpha$ is the fine structure constant and $v_e$ is the velocity
of the electron in the center-of-momentum frame.

The neutrino energy is measured in the laboratory frame, not in the
center-of-momentum frame.  Therefore, the rate of
production of neutrinos of definite observed energies, $q_{\rm obs}$,
must be computed.
The neutrinos that are observed experience a Doppler shift because of
the motion, $\roarrow{V}$, of the center-of-momentum frame relative to the
laboratory frame.  Let the $z$ axis be
oriented along the direction between the terrestrial detector and the
core of the sun.  Then the non-relativistic expression for the Doppler
shift is
\begin{equation}
q_{\rm obs} = q_{c.m.}(p) \left(1 - V_{z, c.m.}\right) ,
\label{nrDoppler}
\end{equation}
which corresponds to a center-of-momentum velocity in the $z$ direction
of
\begin{equation}
V_{z,c.m.} = \frac{P_{z}}{M} =
\frac{q_{c.m.}(p) - q_{\rm obs}}{q_{c.m.}(p)} .
\label{cmmomentum}
\end{equation}
It is also convenient to introduce the relative energy in the
center-of-momentum frame, i.e.,

\begin{equation}
E =
\frac{p^2}{2 \mu} .
\label{cmenergy}
\end{equation}
The observed energy of the neutrino depends upon the direction in which
the decaying nucleus is moving [via Eq.\ (\ref{nrDoppler})], but only on the
magnitude of the relative energy (not on the direction of $\roarrow{p}$),
i.e.,

\begin{equation}
q_{c.m.}(p) = q_{\rm cont,star} + p^2/2\mu
= q_{\rm cont,star} + E.
\label{qcmversusE}
\end{equation}
Carrying out the integrations over the unimportant directions and
converting from an integration over relative momentum to
an integration over relative energy, one obtains
\begin{eqnarray}
d\ {\rm Rate} (T) &&= C^\prime(T) \int^\infty_0
dE\int^{+\infty}_{-\infty}
dP_z \exp
(-E/T)\exp (-P^2_z/2MkT)\nonumber\\
&&\times q^2_{c.m.}(E)
\left[1 + q/M\left(^7{\rm Li}\right)\right]^{-1}\left[1 - \exp
\left(-8\pi \alpha m_e/p\right)\right]^{-1}.
\label{contbasicrelation}
\end{eqnarray}

The most important terms in Eq.\ (\ref{contbasicrelation}) have direct physical
interpretations.  The function $\exp(-E/T)$ represents the
Maxwell-Boltzman distribution of relative internal energies, $E$, in the
center-of-momentum frame and is dominated by the
electron kinetic energy.
The function $ \exp (-P^2_z/2MkT) $ describes the Maxwell-Boltzman
distribution of the center-of-momentum velocities and is
responsible for the
Doppler shifts (via Eq.\ \ref{nrDoppler}).
Since the ${\rm ^7Be}$ nucleus is much heavier than an electron, the
center-of-momentum frame approximately coincides with the rest frame of
the  capturing ${\rm ^7Be}$ ion.
Thus the velocities of the ${\rm ^7Be}$ nuclei
essentially determine the Doppler shifts.
The last three terms in Eq.\ (\ref{contbasicrelation})
represent, respectively, the neutrino phase space ($q^2$), a small
correction resulting from conservation of energy that is associated with
the recoil energy of the ${\rm ^7Li}$ nucleus, and a small correction to the
basically $1/v$-dependence [see Eq.\ (\ref{xinonrelativistic})]
of the probability density of the electron near the nucleus.

The exponent that describes the Doppler shift is
\begin{equation}
\frac{P^2_z}{2 M kT}\  =\
\frac{M \left( q_{\rm obs} - q_{c.m.}(E)\right)^2}{2 kT q^2_{c.m.} (E)}
,
\label{Dopplerexponent}
\end{equation}
which can be written in a convenient numerical form as
\begin{equation}
\frac{P^2_z}{2 M kT}\  =\
\frac{51.023}{T_6} \frac{\left(q_{\rm obs} - q_{\rm cont,star}
- E \right)^2}{\left(1 + E/q_{\rm cont,star}\right)^2} ,
\label{Dopplernumerical}
\end{equation}
where $T_6$ is the local temperature in the sun measured in units of
$10^6$ K and the neutrino energy and the internal kinetic energy are
expressed in keV.
The large numerical coefficient in the exponent that is shown in Eq.\
(\ref{Dopplernumerical}) forces $q_{\rm obs}$ to be equal to
$q_{\rm cont,star} +  E $ to within a fraction of a keV.
Expressed in terms of the observed neutrino energy, Eq.\
(\ref{contbasicrelation}) can be rewritten as:
\begin{eqnarray}
\frac{d\ {\rm Rate} \left(T, q_{\rm obs}\right)}{d q_{\rm obs}}
&&= C^\prime(T) \int^\infty_0
dE\  \exp (-E/T)
\exp\left(\frac{-51.023}{T_6} \frac{\left(q_{\rm obs} - q_{\rm cont,star}
- E \right)^2}{\left(1 + E/q_{\rm cont,star}\right)^2}\right)
\nonumber\\
&&\times q^2_{c.m.}(E)
\left[1 + q/M\left(^7{\rm Li}\right)\right]^{-1}\left[1 - \exp
\left(-8\pi \alpha m_e/p\right)\right]^{-1},
\label{rebasicrelation}
\end{eqnarray}
which is the principal result of this section.

The normalized
spectrum of neutrino energies due to electron capture from continuum
orbits within a spherical shell at temperature $T$ is
\begin{equation}
{\rm Spectrum}_{\rm cont}\left(T,q_{\rm obs}\right) = \frac{d\ {\rm
Rate}\left(T, q_{\rm obs}\right)/d q_{\rm obs}}{\sum_{q_{\rm obs}}d\ {\rm
Rate}\left(T, q_{\rm obs}\right)/d q_{\rm obs}} .
\label{dratenormalized}
\end{equation}
The average over spherical shells at different temperatures will be
described in Sec.\ \ref{sec:energyprofile}, where use will be
made of detailed solar models.

\section{Capture from Bound Electron Orbits}
\label{sec:boundcapture}

A small, but significant fraction of electron captures in the sun
occur from bound orbits. This section describes the
calculation of the neutrino energy profile for
bound electron captures.
The rate of bound-capture in the sun was
first evaluated by Iben and his collaborators\cite{Iben67} in an
elegant paper which carried out a variational-principle calculation of
the binding energies and eigenfunctions in the presence of the solar plasma
and which also presented
formulae for the fractional occupation of different bound
atomic levels as a function of the ambient physical variables.

The normalized energy profile resulting from bound capture has the
simple form
\begin{equation}
{\rm Spectrum}_{\rm bound}\left(q_{\rm obs},T\right) =  q_{\rm
bound,star}^{-1}\left(\frac{M({\rm ^7Be})c^2}{2\pi kT}\right)^{1/2}
\exp\left[ - \frac{M({\rm ^7Be})c^2}{2 kT}\left(\frac{q_{\rm obs}
 - q_{\rm bound,star}}
{q_{\rm bound,star}}\right)^2\right],
\label{boundenergyprofile}
\end{equation}
where ${q_{\rm bound,star}}$ is given in Eq.\ (\ref{qboundstargs})
and Eq.\ (\ref{qboundstares}).
The spectrum given by Eq.\ (\ref{boundenergyprofile}) is obtained
directly from the Maxwell-Boltzman distribution for the ${\rm ^7Be}$ ions by
substituting for $P_z$ the expression given in Eq.\ (\ref{cmmomentum})
for the ion momentum.
For the capture of bound electrons,
the center-of-momentum frame coincides with the rest frame of the
${\rm ^7Be}$ ion, which is why the kinematic complications that are present
for the case of continuum capture (in which
both electrons and ${\rm ^7Be}$ ions have non-zero
translational velocities) are absent for bound capture.

The fraction, $f_{\rm bound}/( 1.0 + f_{\rm bound})$,
of electron captures that occur from bound K-shell orbits of ${\rm ^7Be}$ at a
fixed temperature is given by the following expression\cite{Iben67}
\begin{equation}
f_{\rm bound}(T) = \left(5.07/T_6\right) S_R \exp \left(2.515
\sigma_R/T_6\right),
\label{fbound}
\end{equation}
where $S_R$ and $\sigma_R$ are quantities that result from the
variational principle calculation.  Bahcall and Moeller\cite{Bahcall69}
have given a convenient formula for $\sigma_R$
[see Eq.\ (\ref{sigmaRdefinition}) for the definition of $\sigma_R$].
They give
\begin{mathletters}
\label{allequations}
\begin{equation}
\sigma_R \cong - 0.431 + 2.091 r - 1.481 r^2 + 0.401 r^3~,
\label{sigmaRinr}
\end{equation}
in terms of the dimensionless Debye-Huckel screening length,~$r$ (the
Debye-Huckel screening length divided by the Bohr radius),
where:
\begin{equation}
r = 0.298 \left[ 64 T_6/\rho \left( 3 + X \right) \right]^{1/2}~,
\label{rdefinition}
\end{equation}
with $\rho$ and $X$ being, respectively, the local density (in
gm~cm$^{-3}$)
and the hydrogen mass fraction.
The quantity $S_R$ can be calculated from formulae given
in\cite{Iben67}, i. e.,
\begin{equation}
S_R = C^2_R D^{-1}\left[1 + 0.435 L_R \exp \left(-0.735
\sigma_R/T_6\right)\right]~,
\label{equationa}
\end{equation}

\begin{equation}
D = \left[1 + L_R + 0.25 L^2_R \exp \left(-0.735
\sigma_R/T_6\right)\right]~,
\label{equationb}
\end{equation}
\begin{equation}L_R = 0.246 \left(\rho\mu_e^{-1}T_6^{-3/2}\right) \exp
\left(2.515 \sigma_R/T_6\right)~,
\label{equationd}
\end{equation}
where $\mu_e$ is the electron mean molecular weight, and\cite{Bahcall69}
\begin{equation}
C^2_R \cong -0.6064 + 4.859 r - 5.283 r^2 + 1.907 r^3~.
\label{equatione}
\end{equation}
\end{mathletters}

The bound enhancement, $f_{\rm
bound}$, averaged over the conditions of a standard solar model yields a
value of $f_{\rm bound} \approx 0.21$\cite{Bahcall69} (see also column 5
of Table 3
of the present paper).

\section{Characteristics of the Energy Profile}
\label{sec:energyprofile}

This section describes the characteristics of the energy
profiles of the two ${\rm ^7Be}$ neutrino lines that are shown in Fig. 1.
The calculated profiles have been averaged over the
physical parameters of detailed solar models.
The discussion makes use of the energy spectra that were
computed for fixed temperatures
in Sec.\ \ref{sec:continuumcapture} for electron captures from
continuum orbits [see especially Eq.\ (\ref{contbasicrelation})
and Eq.\ (\ref{dratenormalized})]
and in Sec.\ \ref{sec:boundcapture} for
captures from bound orbits
[see especially
Eq.\ (\ref{boundenergyprofile}) and Eq.\ (\ref{fbound})].

The energy spectra are given numerically in Tables I and II.
The most striking characteristic of the line profiles is their
asymmetry. Figures 2 and 3 display the calculated line profiles.
The characteristic shapes, above and below the energy at
which the probability for neutrino emission is a maximum, will be explained
physically and derived analytically (approximately) in the following
section, Sec.\ \ref{sec:approximate}.

The basic properties of the line profile
that are computed numerically in this section
are the shift in average neutrino energy, $\Delta$, relative to the
laboratory energy of the neutrino line, the full-width-at-half-maximum
(FWHM) of the line profile, the half-width at half-maximum
for energies above $(W_+)$ and below $(W_-)$ the peak of
the line profile, and the first, second, and third moments of the energy
distributions.
For all the solar models that are described in Table III,
these characteristics of the energy profile are given
in Table V for both ground-state capture (862 keV,
see Fig. 1) and excited-state capture (384 keV, see
Fig. 1).

\subsection{Solar Models}
\label{sec:solarmodels}

The principal characteristics of the
four solar models\cite{Bahcall88,Bahcall92,Bahcall82}
that were used to compute the averaged line profiles are
summarized in Table III.
The preferred solar model, which is listed first in both Table III and
Table V, was computed by Bahcall and Pinsonneault
\cite{Bahcall92} and is the only detailed solar interior model published
to date
that includes helium diffusion. For comparison, I calculate the line
profile
with a model labeled (No diffusion) that was computed\cite{Bahcall92}
with the same
input data that was used in calculating the preferred model, except that
the ``No diffusion'' model does not include helium diffusion.
As a test of the robustness of the results,
I have also calculated the energy profiles with the aid of standard
solar models
computed earlier, in 1988
\cite{Bahcall88} and in 1982\cite{Bahcall82}, which used
less accurate input data.  Note that the number of spherical shells
used in the interior region in which ${\rm ^7Be}$ neutrinos are produced
increases monotonically from 19 shells used in the 1982 solar model to 127
shells used in the 1992 models.  The results given below show that there
are no significant differences among the line profiles computed with the
three high-resolution models (with interior shells $\geq 87$).

The solar models listed in Table~III span a decade of state-of-the-art solar
research. All four models have
the same central temperature to within
$\pm 0.5$\%,
\begin{equation}
T_c = (15.58 \pm 0.08)
\times 10^6~{\rm K} .
\label{Tsubc}
\end{equation}
The central temperature for the three more-precise
models varies over a range of \hbox{$\pm 0.3\%$.}
The less accurate 1982 model has a central
temperature that differs by 0.8\% from the average modern value.

The average temperature, $<T>_{^7\rm Be}$,
computed by weighting
the temperature in each spherical shell  according to the
${\rm ^7Be}$ neutrino flux produced in that shell,

\begin{equation}
<T>_{^7\rm Be} = \sum_T d \phi\left(^7{\rm Be}, T\right) {\rm T},
\label{TsubBe7}
\end{equation}
is the same to within
$\pm 0.3$\% for the three solar models from 1988 and 1992, i.e.,
\begin{equation}
\langle T\rangle_{^7{\rm Be}} = (14.10 \pm 0.04) \times 10^6\ {\rm K}.
\label{TBenumber}
\end{equation}
The 1982 model has a ${\rm ^7Be}$-weighted temperature of
$\langle T\rangle_{ ^7{\rm Be}} = 13.6 \times
10^6~{\rm K}$, which differs by
about 3.5\% from the more accurate later models,
presumably because of the small number of shells
(\ref{solaratomicbinding}) (used in
the earlier calculation.  The variation over the past decade in the computed
total ${\rm ^7Be}$ neutrino flux is $\pm 6$\% ;
the decadal variation in the computed $^8$B neutrino flux is also
$\pm 6$\%.

The relative number of electron captures that occur
from bound orbits,$f_{\rm bound}$,
is robustly determined by the
solar models.  From Table III,

\begin{equation}
f_{\rm bound} = 0.221 \pm 0.006 .
\label{numericalfbound}
\end{equation}
This result is in good agreement with the 1969 value
\cite{Bahcall69} of $f_{\rm bound} = 0.21$.

Table~IV lists the central temperatures for ten other
recently-published solar models that were calculated by different groups
using different computer codes and different input
parameters\cite{Berthomieu93}, as well as the 1988 and 1992
(no-diffusion) solar models described in Table~III.
None of the
solar models listed in Table~IV includes helium diffusion.
The heterogeneous set of
models referred to in Table~IV were
derived for a variety of different applications, most of which were not
directly related to solar neutrinos.  The applications often did not
require the highest-obtainable accuracy for the solar interior
conditions.  The precision with which the models were constructed and
the accuracy of the input data varies from model to model.  In many
cases, the authors did not use the best-available radiative opacities and
nuclear reaction rates.   The central
temperatures given by the variety of solar models listed in Table~IV can
be summarized by the relation
\begin{equation}
T_{\rm central} ({\rm no\ diffusion}) = 15.55 (1 \pm 0.01) \times 10^6~{\rm K}
{}.
\label{Tempnodiffusion}
\end{equation}
All of the
central temperatures in Table~IV
would be increased by $\simeq 0.1 \times 10^6$~K if
diffusion were included.

Taken together, the results shown in Table~III and Table~IV demonstrate
that the central temperature of the sun is determined to $\pm 1\%$ even
without requiring unusual precision and accuracy in the calculations.
In what follows, I shall only make use of the four solar models listed
in Table~III, since these are the only models, with which I am
acquainted, whose characteristics are published in sufficient detail to
permit precise calculations of the $^7$Be line profile.

\subsection{Properties of the Energy Spectrum}
\label{sec:energyspectrum}

The neutrino energy spectrum that is produced in a spherical
shell of the sun that is at a specified temperature, $T$,
is obtained by adding the
normalized energy spectra for the capture of electrons from
continuum orbits (computed in Sec.\ \ref{sec:continuumcapture})
and the normalized spectrum
from bound orbits (computed in Sec.\ \ref{sec:boundcapture}).
The relative contributions are weighted by the factor
$f_{\rm bound}$ that was defined in Eq.\ (\ref{fbound}).
Thus
\begin{equation}
{\rm Spectrum} \left(T,q_{\rm obs}\right) = \frac{{\rm Spectrum}_{\rm
cont} \left(T, q_{\rm obs}\right) + f_{\rm bound} (T) {\rm
Spectrum}_{\rm bound}
\left(T, q_{\rm obs}\right)}{1 + f_{\rm bound} (T)} .
\label{spectrumcombined}
\end{equation}
The neutrino spectrum for the entire sun predicted by a particular
solar model is the weighted average
of ${\rm Spectrum}\left(T, q_{\rm obs}\right)$,
weighted with respect to the $^7$Be flux that is produced at each
temperature.  Therefore, the neutrino spectrum predicted by a given solar
model is

\begin{equation}
{\rm Spectrum}_{\rm solar}\left(q_{\rm obs}\right) = \sum_T d
\phi\left(^7{\rm Be}, T\right) {\rm Spectrum}\left(T, q_{\rm
obs}\right),
\label{averagespectrum}
\end{equation}
where the $^7$Be neutrino flux is normalized to unity when integrated
over the whole star,
\begin{equation}
\sum_T d \phi \left(^7{\rm Be}, T\right) = 1.000 .
\label{normalized7Be}
\end{equation}
The values of the weighting factors, $d \phi \left(^7{\rm Be},
T\right)$, are given in published tables\cite{Bahcall88,Bahcall92,Bahcall82}
for all four of the models used here.  The numerical values for ${\rm
Spectrum}_{\rm solar}(q_{\rm obs})$ are given in Tables I and II.
%Figures 1 and 2 go here.
%\input table1.tex
%\input table2.tex

The predicted probability distribution for the neutrino energy
spectrum has a maximum at a
well-defined energy, $q_{\rm peak}$, whose location
is obvious in Fig. 2 and in Fig. 3.
The peak of the probability distribution exceeds by a small amount,
$\delta q$ [see Eq.\ (\ref{energycontstar}) and Eq.\
(\ref{contstarvalue})],
 the neutrino energy, $q_{\rm cont,star}$, that
corresponds to capturing an electron from a continuum orbit with zero
kinetic energy , i. e.,
\begin{equation}
\delta q\  \equiv\  q_{\rm peak} - q_{\rm cont,star}.
\label{deltaqdefinition}
\end{equation}
For the most accurate available standard solar model, which is
labeled ``Bahcall-Pinsonneault 1992 (Helium diffusion)'' in Table III and
Table V, the shift $\delta q$ for the (more energetic)
ground-state transition is

\begin{equation}
\delta q~{\rm (g.s.)} = 0.23~{\rm keV} ,
\label{deltaqgs}
\end{equation}
and is
\begin{equation}
\delta q~{\rm (ex.s.)} =
0.04~{\rm keV}\
\label{deltaqexs}
\end{equation}
for the (less energetic) excited-state transition.
Essentially identical values of $\delta q$ are obtained with the other
solar models used here (see the last column of Table V).
The value of $q_{\rm peak}$ can be calculated using the last column of
Table V and the relation given in Eq.\ (\ref{deltaqdefinition}), i.e.,
\begin{equation}
q_{\rm peak} = q_{\rm cont, star} + \delta q .
\label{qpeakdefinition}
\end{equation}
The value of $q_{\rm peak}$ for ground-state transitions is

\begin{equation}
q_{\rm peak}~{\rm (g.s.)} = \left(862.27 \pm 0.01\right)~{\rm keV}~,
\label{gsqpeak}
\end{equation}
0.43~keV larger than the laboratory decay energy (see Sec.\
\ref{sec:labdecays}). For excited-state decays, the energy peak occurs at
\begin{equation}
q_{\rm peak}~{\rm (g.s.)} = \left(384.47 \pm 0.01\right)~{\rm keV} ,
\label{exsqpeak}
\end{equation}
0.19~keV larger than the laboratory decay energy (Sec.\
\ref{sec:labdecays}).

It is convenient to define the n-th moment of the solar energy spectrum
by the relation

\begin{equation}
\langle\left(q - q_{\rm peak}\right)^n \rangle \equiv \int^\infty_0 d q
\,{\rm Spectrum}_{\rm solar}(q) \left(q -
q_{\rm peak}\right)^n .
\label{nthmoment}
\end{equation}
The moments are computed about the energy $q_{\rm
peak}$ at which the probability
distribution peaks.

The shift in average neutrino energy,
$\Delta$, from the laboratory value to the
solar interior value is

\begin{equation}
\Delta\ \equiv \Bigl\langle q - q_{\rm lab}\Bigr\rangle ,
\label{Deltadefinition}
\end{equation}
\noindent
or, using the definition of the spectrum moments given in Eq.\
(\ref{nthmoment}),
\begin{equation}
\Delta\ = \Bigl\langle q - q_{\rm peak}\Bigr\rangle +
\left(q_{\rm peak} - q_{\rm cont,star}\right) +
\left(q_{\rm cont,star} - q_{\rm Lab}\right) .
\label{ReDeltadefinition}
\end{equation}
The first two terms in Eq.\ (\ref{ReDeltadefinition}) are given in
columns 3 and 9 of Table V.  The last term can be computed
from Eq.\ (\ref{gsenergylab}), Eq.\ (\ref{esenergylab}),
Eq.\ (\ref{contstarvalue}) and Eq.\ (\ref{qcontes}) of Sec.\
\ref{sec:energies}.
The value of $\left(q_{\rm cont,star} - q_{\rm Lab}\right)$ is 0.20 keV
for ground-state transition and 0.15 keV for
excited-state transition.
Combining the results given in Table V,
the shift
for the ground-state transition is, for all three of the
modern (1988--1992) models,
\begin{equation}
\Delta~{\rm (g.s.)} = 1.29~{\rm keV},
\label{Deltagsgs}
\end{equation}
\noindent
with a spread of only $\pm 0.5\%$.  The earlier (1982) model yields
$\Delta {\rm (g.s.)} = 1.23\ {\rm keV}$.
The shift for the excited-state transition is
\begin{equation}
\Delta~{\rm (ex.s.)} = 1.24~{\rm
keV}
\label{Deltagses}
\end{equation}
for the three modern models with a spread of only $\pm 0.5\%$.  The
earlier (1982) model gives a 4\% smaller value, 1.18~keV.

Table V also presents some of the other calculated characteristics of the
energy profile.  The full-width-at-half maximum of the profile is
denoted by $FWHM$ and is listed in column 4 of Table V; the half-width on
the low-energy side of the peak (column 5) is denoted by $W_{-}$ and the
half-width on the high-energy side (column 6) is denoted by $W_{+}$.
The most accurate values for the ground-state transition
are
\begin{equation}
{\rm FWHM}~{\rm (g.s.)} = 1.63~{\rm keV}\ ;\ W_- {\rm (g.s.)}
 = 0.56~{\rm keV}\ ;\ W_+ {\rm (g.s.)} = 1.07~{\rm keV}.
\label{HMgs}
\end{equation}
The corresponding values for the excited-state transition
are
\begin{equation}
{\rm FWHM}~{\rm (ex.s.)} = 0.97~{\rm keV}\ ;\ W_- {\rm (ex.s.)}
= 0.24~{\rm keV}\ ;\ W_+ {\rm (ex.s.)} = 0.73~{\rm keV}\ .
\label{HMes}
\end{equation}

Figures 2 and 3 are remarkably asymmetric.  The degree of asymmetry can
be quantified by taking the appropriate dimensionless ratio of the
third and the second moments that is known as the skewness.  For our
case, the skewness of the profile is given by
\begin{equation}
{\rm Skewness} \equiv \frac{\langle \left(q - q_{\rm peak}\right)^3
\rangle^2}{\langle\left(q - q_{\rm peak}\right)^2\rangle^3} .
\label{skewnessdefinition}
\end{equation}
The values of the skewness that are computed from columns 7 and 8 of
Table V are, respectively,
\begin{equation}
{\rm Skewness}~{\rm (g.s.)} = 5.7\ ;\ {\rm Skewness}~{\rm (ex.s.)} = 5.5\ .
\label{Skewnessvalues}
\end{equation}
The skewness of any symmetric distribution, such as the normal
distribution, is equal to zero.

The shape of the energy profile can be well-described by simple analytic
functions.  For neutrino energies less than the peak energy, the profile
is essentially Gaussian, i. e.,
\begin{equation}
{\rm Spectrum}_{\rm solar}\left(q_{\rm obs}\right)\ \cong\
N \exp\left[-\left(q_{\rm obs} - q_{\rm peak}\right)^2/2 w^2\right]\
,\  q_{\rm obs} < q_{\rm peak} .
\label{Gaussianspectrum}
\end{equation}
The high-energy tail is well described by a Boltzmann distribution, i. e.,

\begin{equation}
{\rm Spectrum}_{\rm solar}\left(q_{\rm obs}\right)\ \cong\
N \exp\left[-\left(q_{\rm obs} - q_{\rm peak}\right)/T_{\rm eff}\right]\
,\ q_{\rm obs} > q_{\rm peak}.
\label{Boltzmannspectrum}
\end{equation}
In Eq.\ (\ref{Gaussianspectrum}) and Eq.\ (\ref{Boltzmannspectrum}), the
quantity $N$ is a normalization factor.
For ground-state capture, the effective width, $w$, to be
used in
Eq.\ (\ref{Gaussianspectrum}) is approximately
\begin{mathletters}
\label{twosigmas}
\begin{equation}
w({\rm g.s.}) \ =\ 0.48\ {\rm keV}\ .
\label{sigmags}
\end{equation}
The corresponding value for excited-state capture is
\begin{equation}
w({\rm ex.s.}) \ =\ 0.20\ {\rm keV}\ .
\label{sigmaexs}
\end{equation}
\end{mathletters}
Both transitions are well-described by a single effective temperature,
\begin{mathletters}
\label{twotemperatures}
\begin{equation}
kT_{\rm eff}\ =\ 1.31\ \pm\ 0.02~{\rm keV}\ ,
\label{temperaturekev}
\end{equation}
i. e.,
\begin{equation}
T_{\rm eff}\ \cong\ 15.1 \times 10^6~{\rm K}.
\label{temperaturek}
\end{equation}
\end{mathletters}
The uncertainty indicated in the value of the effective temperature,
$T_{\rm eff}$, reflects the fact that Eq.\ (\ref{Boltzmannspectrum})
is only approximate and the best-fitting value of $T_{\rm eff}$ varies
somewhat with neutrino energy.

\section{Approximate Analytic Derivations of the Energy Shift and Energy
Profile}
\label{sec:approximate}

This section gives approximate analytic derivations of the
principal characteristics of the neutrino energy spectrum from
$^7$Be electron capture in the sun.  The purpose
of this discussion is to provide insight regarding the physical origins
of the profiles that are shown in Fig. 2 and Fig. 3. For simplicity, I
will usually concentrate on the profile created at a
fixed temperature, $T_{\rm 6}$, ignoring
the average over the temperature distribution of the center of the sun.
In some cases, I will also ignore
the small (20\%)
contribution to the energy profile that arises from
captures from bound states (cf. discussion in Sec.\
\ref{sec:boundcapture}).
In all cases,
I will omit small terms of order $q/M(^7{\rm Li})$,
the ratio of the neutrino energy, $q$,
to the mass of the $^7$Li nucleus, and
terms of order,$E/q$, the ratio of the relative
kinetic energy,$E$, of the electron and of the nucleus
to the neutrino energy.

These approximations permit the isolation of the principal physical
processes that determine the neutrino energy profile
and allow analytic calculations to be made that
reproduce the general features of the detailed numerical results.

I begin by deriving approximate expressions for the energy half-widths,
$W_-$ and $W_+$,
and then obtain an analytic expression for the energy shift, $\Delta$.

\subsection{Energy Half-Widths}
\label{sec:energyhalf}

The low-energy side of the profile,
$q_{\rm obs} < q_{\rm peak}$,
is produced by $^7$Be nuclei that are moving away from the observer
and which capture electrons with small or zero kinetic
energies (i. e., low-velocity electrons from the continuum
or electrons in bound orbits).
For such captures, the reaction rate is given approximately by
[see Eq.\ (\ref{contbasicrelation}), Eq.\ (\ref{Dopplernumerical}),
and Eq.\ (\ref{boundenergyprofile})]
\begin{equation}
d~{\rm Rate}~\stackrel{\textstyle \propto}{\sim}
{}~{\rm const} \times \exp \left[- \frac{51}{T_6} \left(q_{\rm obs} -
q_{\rm cont,star} - \delta q \right)^2\right].
\label{dRatesmallside}
\end{equation}
Replacing in Eq.\ (\ref{dRatesmallside}) the
expression $\left(q_{\rm obs} -
q_{\rm cont,star} - \delta q \right)$ by  $W_{-}$,
the half-width-at-half maximum
on the low-energy side, one has
\begin{equation}
W_- \approx \left(\ln 2 \langle T_6\rangle/51\right)^{1/2}.
\label{Wminusexpression}
\end{equation}
Evaluating Eq.\ (\ref{Wminusexpression}) for a characteristic $^7$Be
temperature of $14 \times 10^6$K (see discussion of solar models in
Sec.\ \ref{sec:solarmodels}) yields
\begin{equation}
W_- \approx 0.44~{\rm keV}.
\label{Wminusapproximation}
\end{equation}
This result is in satisfactory agreement with the low-energy width that
is calculated numerically, $W_{-} = 0.56$ keV (see Table V).

The high-energy side of the profile,
$q_{\rm obs} >  q_{\rm peak}$,
is produced by electrons and $^7$Be nuclei with significant internal
kinetic energies,$E$ [see Eq.\ (\ref{contbasicrelation})]. Since the
product of the separate Boltzmann distributions for the electrons and for
the $^7$Be ions is a Boltzmann distribution in the center-of-momentum
frame, the high-energy part of the profile has the approximate shape
\begin{equation}
d~{\rm Rate}~\stackrel{\textstyle \propto}{\sim}~{\rm const} \times
\exp ( - E/T)\ ,
\label{Wplusapproximation}
\end{equation}
where
\begin{equation}
E = p^2/2\mu ,
\label{Eofp}
\end{equation}
and the relative momentum, $p$, is defined by Eq.\ (\ref{definitionp}).
One can define an effective temperature, $T_{\rm eff}$, for the
high-energy tail of the spectrum by fitting at two energies
the expression given in
Eq.\ (\ref{Wplusapproximation}) to the numerically-computed energy
spectra that are given in Table I and Table II (and shown in Fig.3 and
Fig.4).
Symbolically,
\begin{equation}
T_{\rm eff} \approx \frac{q_{\rm obs}{(2)} - q_{\rm obs}{(1)}}{\ln
\left[{\rm Spectrum_{obs}}{\left(q_{\rm obs}(1)\right)}/{\rm Spectrum_{obs}}
{\left(q_{\rm obs}(2)\right)}\right]} ,
\label{Teffdefinition}
\end{equation}
where $q_{\rm obs}(1)$, $q_{\rm obs}(2)$ are typical energies in the
line profile.  The value of $T_{\rm eff}$ depends slightly on the choice
of $q_{\rm obs}(1)$, $q_{\rm obs}(2)$.
For both transitions shown in Fig. 1,

\begin{equation}
T_{\rm eff} \cong 1.31~{\rm keV} \approx 15 \times 10^6~{\rm K},
\label{Teffvalue}
\end{equation}
which translates into a half-width-at-half-maximum of

\begin{equation}
W_+ \cong (\ln 2) T_{\rm eff} \approx 0.91~{\rm keV} .
\label{Wplusvalue}
\end{equation}
The value for $W_+$ given in Eq.\ (\ref{Wplusvalue}) is approximately
halfway between the accurate values given in Table V
for the ground-state decay (1.07 keV)
and the excited-state decay (0.73 keV).

The reason that the effective temperature given in Eq.\ (\ref{Teffvalue})
is slightly larger than the value obtained from the solar model and
given in Eq.\ (\ref{TBenumber}) is that Eq.\ (\ref{Teffdefinition})
is only approximate.  A more accurate (but less transparent)
determination of $T_{\rm eff}$
can be obtained by taking the ratio of integrand in
Eq.\ (\ref{rebasicrelation}) at two neutrino energies and setting that
ratio equal to the ratio of the spectrum probabilities.
This more accurate relation should be used in interpreting future
experiments.
\subsection{The Energy Shift: $\Delta$}
\label{sec:energyshift}

The energy shift, $\Delta$, can be written as the sum of two terms,
$\langle q_{\rm obs} - q_{\rm cont,star}\rangle$
plus $(q_{\rm cont, star} - q_{\rm lab})$
[cf. Eq.\ (\ref{ReDeltadefinition})]. For the ground-state transition,
$(q_{\rm cont, star} - q_{\rm lab})$ is 0.20 keV and is
0.15 keV for the excited state transition [see discussion in Sec.\
\ref{sec:energies} and
in Sec.\ \ref{sec:energyspectrum} following Eq.\
(\ref{ReDeltadefinition})].
The dominant term in
the energy shift, $\langle q_{\rm obs} - q_{\rm cont,star}\rangle$,
is equal (to numerical accuracy, which is better than 1\%)
to the same value,
$ 1.09$ keV, for both ground-state and excited-state transitions.
This numerical equality can be established as follows.
The quantity
$\langle q_{\rm obs} - q_{\rm cont,star}\rangle$
is equal to the sum of two terms,
$\langle q_{\rm obs} - q_{\rm peak}\rangle + \delta q$,
both of which are given in Table V.
Adding the two terms in Table V gives, for all of the solar models,
essentially identical values for
$\langle q_{\rm obs} - q_{\rm cont,star}\rangle$.

At first sight, it is surprising that the energy shift relative
to $q_{\rm cont,star}$,
$\langle q_{\rm obs}~-~q_{\rm cont,star}\rangle$,
is independent of the value of $q_{\rm cont,star}$.
However, a simple analytic argument, given below, shows  that, at a
specified temperature $T$, the shift
$\langle q_{\rm obs} - q_{\rm cont,star}\rangle_{\rm T}$ is, to an
excellent approximation, the mean kinetic energy of the electron
and the $^7$Be nucleus that interact to produce the electron capture
reaction.  The mean kinetic energy is, of course, independent of
$q_{\rm cont,star}$.

The average value of the shift relative to $q_{\rm cont,star}$ is
\begin{equation}
\langle q_{\rm obs} - q_{\rm cont,star}\rangle_T = \int^\infty_0 d
q_{\rm obs}\left(q_{\rm obs} - q_{\rm cont,star}\right)\ {\rm
Spectrum}_T (q_{\rm obs}),
\label{Shiftqcontstar}
\end{equation}
which can be rewritten approximately using Eq.\ (\ref{contbasicrelation})
and Eq.\ (\ref{Dopplerexponent}) as

\begin{eqnarray}
\langle q_{\rm obs} - q_{\rm cont,star}\rangle_T &&\ \cong\  T^{-1}
\left(\frac{\beta \pi}{q^2_{\rm cont,star}}\right)^{1/2}\int^\infty_0 d
E\ \exp (-E/T)\nonumber\\
&&\times \int^{+\infty}_{-\infty} dq_{\rm obs} \left(q_{\rm obs} - q_{\rm
cont,star}\right)\ \exp\left[- \beta \left(q_{\rm obs} - q_{\rm
cont,star} - E\right)^2/q^2_{\rm cont,star}\right].
\label{Shiftintegral}
\end{eqnarray}
The key fact that simplifies Eq.\ (\ref{Shiftintegral}) and
results in the equality of the energies for ground-state and
excited-state transitions is that the quantity, $\beta$, which appears in
the second exponent in Eq.\ (\ref{Shiftintegral}), is very large.
Eq.\ (\ref{Dopplerexponent}) and Eq.\ (\ref{Dopplernumerical})
show that $\beta$ is just the ratio of the nuclear rest mass energy
($\sim 7 $ GeV) to the ambient temperature ($\sim 1$ keV), i. e.,
\begin{equation}
\beta = \frac{Mc^2}{2kT}\  \gg\  1 .
\label{betadefinition}
\end{equation}
In the limit of large $\beta$,
\begin{equation}
(\beta/\pi)^{1/2}\ \exp\left(-\beta x^2\right)
\hbox
to0pt{$\longrightarrow$\hss}\lower10pt\hbox{$\scriptstyle
\beta \to\infty$} \delta (x),
\label{deltafunction}
\end{equation}
which greatly simplifies Eq.\ (\ref{Shiftintegral}).
The effect of the delta function is to enforce energy conservation
independent of the Doppler shifts, i. e.,
\begin{equation}
q_{\rm obs} - q_{\rm cont,star} = E .
\label{deltaenergyconservation}
\end{equation}
Carrying out the integration over the delta function gives
\begin{equation}
\langle q_{\rm obs} - q_{\rm cont,star}\rangle_T \ \cong\  T^{-1}
\int^{\infty}_0 dEE\ \exp\ (-E/T) = T
\label{Tresult}
\end{equation}
Therefore,
\begin{equation}
\langle q_{\rm obs} - q_{\rm cont,star}\rangle_\odot
\ \cong\
\int_{\odot} d TT d \phi\left(^7{\rm Be},T\right)
{}~=~1.2\  {\rm keV} ,
\label{finale}
\end{equation}
which is, as promised, independent of $q_{\rm cont, star}$.

\section{Other Effects}
\label{sec:othereffects}

This section shows that the effects on the neutrino energy profile
of electrostatic screening energy (Sec.\ \ref{sec:electroenergy}),
of the gravitational redshift (Sec.\ \ref{sec:gravredshift}),
and of collisional broadening (Sec.\ \ref{sec:incoherence})
are much smaller than the dominant terms (calculated in
Sec.\ \ref{sec:continuumcapture}--Sec.\ \ref{sec:energyprofile})
that arise directly from the Maxwellian energy distributions
of the electrons and of the ions.

\subsection{Electrostatic Screening Energy}
\label{sec:electroenergy}

There is a difference in electrostatic screening energy between the initial and
final states of $^7$Be continuum electron capture.  In the initial
state, a $^7$Be nucleus (charge $z = 4 e$)
and a nearby electron jointly interact
electrostatically with a surrounding charge cloud of electrons and ions
that screen the nucleus.  After the electron is captured, only the
nuclear Coulomb potential corresponding to a charge $z = 3 e$ interacts
with the screening cloud.  The difference in electrostatic screening
energy is contributed partly to the energy of the emitted neutrino.

The electrostatic screening
energy, $\delta E_{\rm Coul}$, can be estimated
by generalizing the familiar Debye-Huckel calculation
of the charge distribution and the potential associated with the monopole
field of a single ion. The more general case
consists of a monopole field (from the capturing ion)
plus a dipole field (from
the ion plus the electron that is about to be captured).
Solving Poisson's equation for the monopole plus dipole fields, one
obtains the potential, $\phi_{\rm before}~(\roarrow{r})$,
before the electron capture occurs,
\begin{equation}
\phi_{\rm before}~ (\roarrow{r}) = z_{\rm net} \frac{\exp (-\kappa r)}{r} +
\frac{p\kappa}{r} \exp (-\kappa r)\left[1 + (\kappa
r)^{-1}\right]\cos\Theta ,
\label{phitotal}
\end{equation}
where $\kappa$ is the inverse of the Debye-Huckel screening
length:

\begin{equation}
\kappa^2 = 4\pi e^2 T^{-1} \Sigma_i z^2_i n_i .
\label{defkappa}
\end{equation}
The first term on the right hand side of
Eq.\ (\ref{phitotal}) is the standard Debye-Huckel potential for a
point charge, $z_{\rm net}$, surrounded by a screening cloud,
where the net (point) charge in the present case
is $+ 3 e$.  The second term  is the dipole solution in which $p$ is the
dipole moment produced by the $^7$Be nucleus and the (to-be-captured)
electron.
In the usual approximation that the screening potential is smaller than
the thermal energy, the charge density surrounding the $^7$Be nucleus
and the electron is
\begin{equation}
\rho_{\rm before}~(\roarrow{r}) = -\frac{\kappa^2}{4\pi}
\phi_{\rm before}~(\roarrow{r}) .
\label{defrho}
\end{equation}

The Coulomb energy both before and after the electron capture
can be evaluated from the relation

\begin{equation}
E_{\rm coul} = \frac{1}{2} \int d^3 \roarrow{r} \rho (\roarrow{r})
\phi (\roarrow{r}) .
\label{Coulombenergy}
\end{equation}
In evaluating the initial Coulomb energy,
the potential and the charge distribution are taken from
Eq.\ (\ref{phitotal}) and Eq.\ (\ref{defkappa}) . One may approximate
the final Coulomb energy by assuming that the density distribution of
the charge cloud is unchanged
immediately after the electron capture occurs, but that the final
potential only includes a pure monopole term (3 e/r)
due to the final nucleus plus the potential due to the screening
charges.
The potential due to the screening charges can be taken from the
following defining relation

\begin{equation}
\phi_{\rm before}~ (\roarrow{r}) \equiv \phi_{\rm screening}~
(\roarrow{r})
+ \frac{3e}{r} +
\frac{p}{r^2}{ \cos\Theta} .
\label{phiscreendef}
\end{equation}

The difference in Coulomb energies, $\delta E_{\rm Coul}$,
obtained in this way is

\begin{equation}
\delta E_{\rm Coul} \approx -\frac{4}{15} \frac{e^2}{R_D}
\left(\frac{\lambdabar_D}{R_D}\right) ,
\label{deltaCoulombenergy}
\end{equation}
where $\lambdabar_{D}$ is the De-Broglie wave length of the electron and
$R_{\rm DH} = \kappa^{-1}$ is the Debye-Huckel screening length.
In obtaining Eq.\ (\ref{deltaCoulombenergy}), I evaluated the dipole moment
by assuming that the electron and the $^7$Be nucleus are separated by
a distance $\lambdabar_{\rm D}$ with the origin at the center of
charge.  This yields
$p = 8e \lambdabar_{\rm D}/5$. Inserting typical solar-interior values
into Eq.\ (\ref{deltaCoulombenergy}), one finds

\begin{equation}
\delta E_{\rm approx.~Coul} ~ \approx ~-0.004 ~~ {\rm keV}.
\label{approxdeltaCoulomb}
\end{equation}

One can obtain a conservative upper limit to the change in
electrostatic screening
energy by assuming that the electron is far away from the nucleus when
the electron is captured.
In this extreme limit, one can approximate the change in Coulomb energy
by

\begin{equation}
E_{\rm coul} = \frac{1}{2} \int d^3 \roarrow{r} \rho (\roarrow{r})_{\rm Z=4}
\left[\phi (\roarrow{r})_{\rm Z=4} - \phi (\roarrow{r})_{\rm Z=3}\right] ,
\label{extremeCoulomb}
\end{equation}
where in writing Eq.\ (\ref{extremeCoulomb}) it was also assumed
that the particles in the
screening cloud around the $^7$Be nucleus do not have time to move
during the electron capture. These assumptions yield

\begin{equation}
\delta E_{\rm extreme~Coul} ~ \approx~ - \frac{e^2}{R_D} \simeq ~ - 0.065
{}~~{\rm keV}.
\label{extremedeltaCoulomb}
\end{equation}

Some fraction, $\epsilon < 1$, of the Coulomb energy difference represented by
Eq.\ (\ref{approxdeltaCoulomb}) or Eq.\ (\ref{extremedeltaCoulomb})
should be added to the neutrino energies calculated in the previous
sections.  This addition is estimated to lie between 0.004 $\epsilon$
keV and $0.065 \epsilon$ keV, which is a rather small effect. However, a
quantum mechanical calculation of the change in electrostatic screening
energy would be of interest.

\subsection{Gravitational Redshift}
\label{sec:gravredshift}

Each neutrino energy is shifted by an amount $\Delta q$,
\begin{equation}
\Delta q ({\rm G.R.}) = \frac{-GM(\leq r)}{rc^2} q ,
\label{Deltaqdefinition}
\end{equation}
due to the general relativistic redshift.  In the region in which the
neutrinos are produced, the mean shift
$\Delta q /q \approx 10^{-5}$\cite{Bahcall91}.
Therefore, the average change in the neutrino energy is only
\begin{equation}
\Delta q ({\rm G.R.}) \approx - 0.009~{\rm keV}.
\label{Deltaqvalue}
\end{equation}
This energy redshift is two orders of magnitude smaller than the dominant
processes that are calculated in Secs.\
\ref{sec:continuumcapture}--\ref{sec:approximate}.  The dispersion in
the gravitational energy shift, which would contribute to the broadening
of the line, is an order of magnitude smaller, $\sim 0.001$~keV.

\subsection{Collisional broadening}
\label{sec:incoherence}

The effect of collisions on the coherence length for neutrino oscillations was
first discussed by Nussinov\cite{Nussinov76} in a beautifully-original
paper(that also quantified the small probabilities for obtaining a
large reduction in the electron-neutrino flux at earth due to vacuum
oscillations).  Loeb\cite{Loeb89} was the first to carry out a detailed
calculation of the collisional broadening of a solar neutrino line
(see also\cite{Krauss85}).  In the discussion below, I follow the
 treatment of Loeb.

Let $P(q)$  be the probability that a neutrino is emitted with
the energy $q$ in the presence of the solar plasma and let $q_0$ be the
energy that would be emitted if the collisional frequency were set equal
to zero.
The probability distribution $P(q)$ has the usual Lorentz shape,
\begin{equation}
P(q) = \frac{\pi^{-1} q_\tau}{\left(q - q_0\right)^2 + q^2_\tau},
\label{LorentzP}
\end{equation}
where
\begin{equation}
q_\tau = (2\pi\tau)^{-1},
\label{defineqtau}
\end{equation}
and $\tau$ is the coherence time for neutrino emission\cite{Loeb89}.
The coherence time denotes the period over which the emitting system
looses its phase due to collisions with the background plasma particles.

Loeb\cite{Loeb89} shows that
the collision time for $^7$Be ions in the solar interior
is about $10^{15} {\rm s^{-1}}$ and that the coherence time, $\tau$, is

\begin{equation}
\tau \approx 5 \times 10^{-17}\ \sec .
\label{tauvalue}
\end{equation}
The width, $q_{\tau}$, is therefore very small,
\begin{equation}
q_\tau  \approx 0.013~{\rm keV},
\label{qwidthvalue}
\end{equation}
and may be neglected in the present context.

\section{New Physics}
\label{sec:newphysics}

The two most popular mechanisms for explaining the solar neutrino
problem via new physics are vacuum neutrino oscillations, first
discussed in this connection by Gribov and Pontecorvo\cite{Gribov69} in an
elegant and epochal paper, and matter-enhanced neutrino oscillations,
the MSW effect, a beautiful
idea discovered by
Wolfenstein\cite{Wolfenstein78} and by Mikheyev and
Smirnov\cite{Mikheyev86}.
In Sec.\ \ref{sec:oscillations}, I present calculations of the effect of vacuum
oscillations on the energy shift, $\Delta$, of the $^7$Be neutrino line
(solar versus laboratory) and in
Sec.\ \ref{sec:mswsolution}, I investigate the effect of
matter-enhanced oscillations.
Finally, in Sec.\ \ref{sec:otherprocesses}, I discuss briefly the effect of
other suggested new physics processes on the energy shift.
These discussions
of potential (new-physics) solutions show that the physical processes
considered would not be expected to change significantly the energy shift
that is calculated by considering only thermal effects in the core of the sun.
The effect of the width of the $^7$Be line on neutrino oscillations was
discussed previously by Pakvasa and Pantaleone\cite{Pakvasa90}.

\subsection{Vacuum Oscillations}
\label{sec:oscillations}

If vacuum oscillations occur, the (energy-dependent) probability that an
electron-type neutrino, $\nu_e$,  that is created in the solar interior
with an energy $q_{\rm obs}$ in the laboratory frame survives as an
electron-type neutrino until it reaches a terrestrial detector
modifies the original solar neutrino energy spectrum.  Thus
\begin{equation}
{\rm Spectrum}_{\nu_e}\left(q_{\rm obs}\right) = {\rm Spectrum}_{\rm
solar}\left(q_{\rm obs}\right)\vert\langle \nu_e, {\rm detect}\vert \nu_e, {\rm
emit}
\rangle\vert^2_{q_{\rm obs}}.
\label{Vacuumrelatione}
\end{equation}
If the electron-type neutrino is primarily coupled to only one other
neutrino type,
 e.g., a muon-type neutrino, $\nu_{\mu}$, then the energy spectrum of
the daughter neutrino is
\begin{equation}
{\rm Spectrum}_{\nu_\mu}\left(q_{\rm obs}\right) = {\rm Spectrum}_{\rm
solar}\left(q_{\rm obs}\right)\left[1 - \vert\langle \nu_e,
{\rm detect}\vert\nu_e, {\rm emit}\rangle\vert^2_{q_{\rm obs}}\right].
\label{Vacuumrelationmu}
\end{equation}
The probability for an electron-type neutrino to change its flavor to a
muon-type neutrino can be written
as\cite{Gribov69,Frautschi69,Bilenky78}
\begin{equation}
\vert\langle\nu_{\mu}, {\rm detect}\vert\nu_e, {\rm emit}\rangle\vert^2_{q_{\rm
obs}} =
1 - \vert\langle \nu_e,
{\rm detect}\vert\nu_e, {\rm emit}\rangle\vert^2_{q_{\rm obs}} =
\sin^2 2\Theta_V \sin^2\phi\left(q_{\rm obs}\right),
\label{transitionprobability}
\end{equation}
where $\Theta_V$ is the vacuum oscillation angle and $\phi(q_{\rm obs})$
depends upon the neutrino energy, upon the differences in the square of
the masses of $\nu_{\mu}$ and $\nu_e$, $\Delta m^2$, the distance,
$D$, between the terrestrial detector and the location in the sun where
the neutrinos are produced.  Numerically,
\begin{equation}
\phi\left(q_{\rm obs}\right) = 1.15 \times 10^{11}\left(\frac{0.86~{\rm
MeV}}{q_{\rm obs}}\right)\left(\frac{\Delta m^2}{{\rm
eV}^2}\right)\left(\frac{R}{1~{\rm A.U.}}\right).
\label{phivacuumvalue}
\end{equation}

As pointed out by Bahcall and Frautschi\cite{Frautschi69}, the survival
probability must be averaged over the neutrino energy profile in order
to calculate the effect of vacuum oscillations on the observed neutrino
event rates.  Fine-tuning of the vacuum oscillation parameters is
required to produce large effects on the observed rates.
If vacuum oscillations reduce the flux of $^8$B
electron-type neutrinos (which have a broad energy profile with a total
width $\sim 10$ MeV)
by a large factor (as required to explain the
difference between the standard model predictions and the observations with
the chlorine and Kamiokande detectors), then the mass difference squared,
$\delta m^2$, must cause $\phi(q_{\rm obs})$ (evaluated at $\approx 8$
MeV) to be a small integer
multiple of $\pi/2$\cite{Frautschi69}.  Thus, if vacuum oscillations are to
explain the solar neutrino problem,

\begin{equation}
\Delta m^2 \left({\rm big\ effect}; ^8{\rm B}\right) \sim
10^{-10.5}~{\rm eV}^2.
\label{vacuummassvalue}
\end{equation}

There have recently been several careful studies \cite{Krastev93} of the
constraints on the vacuum oscillation parameters that are implied by the
existing four solar neutrino experiments
\cite{Davis89,Hirata91,Anselmann92,Abazov91}.  Krastev and
Petcov\cite{Krastev93} summarize these results as follows:
\begin{equation}
5 \times 10^{-11}~{\rm eV}^2 \leq \delta m^2 \leq 11.1 \times
10^{-11}~{\rm eV}^2 ,
\label{alloweddeltam2}
\end{equation}
\begin{equation}
0.75 \leq \sin^2 2\Theta_V \leq 1.0 .
\label{allowedsinsquared}
\end{equation}
Only certain combinations of $\delta m^2$ and $\sin^2 2\Theta_V$ are
allowed, but for convenience and to be conservative, I have explored the
entire range in the rectangular space defined by
Eq. (\ref{alloweddeltam2}) and
Eq. (\ref{allowedsinsquared}).  The allowed range of $\phi \left(q_{\rm obs} =
q_{\rm peak}\right)$ is

\begin{equation}
3.5 \times \pi/2 \leq \phi (862.27~{\rm keV}) \leq 8.0 \times \pi/2 .
\label{allowedphi}
\end{equation}

The survival probability for an electron-type neutrino does not change
much over the 1 kev width of the $^7$Be line.
In order for the phase-angle, $\phi(q_{\rm obs})$
(defined by Eq.\ (\ref{phivacuumvalue}) to change by an appreciable
fraction of a  radian,
the observed energy must change by an amount, $\Delta q_{\rm obs}$,
that is much greater than the full width of the neutrino line.
The line profile is
not significantly affected by vacuum oscillations
unless the
phase-angle, $\phi$, is chosen
just so as to make the electron-neutrinos at the peak of the energy
$^7$Be energy
profile maximally mix into muon neutrinos, i. e., $\phi(q_{\rm peak})$
is chosen to be an odd integer
multiple of $\pi/2$.
Even in the case of maximum mixing, for which electron neutrinos are
practically all flavor-converted,  the resulting muon neutrinos have
essentially the same energy spectrum as the original spectrum
with which the electron
neutrinos are created.  Therefore,
the energy profile that would be detected by neutrino-electron scattering
is essentially the same for maximum mixing and for no mixing.
The observed rate is decreased, of course,  by mixing
because muon neutrinos scatter off
electrons about a factor of five less strongly than electron neutrinos.

How can we quantify the effect of vacuum oscillations on the $^7$Be neutrino
line profile?  The most direct effect of oscillations on the $^7$Be line
profile is manifested in the difference, $\Delta$
Eq.\ (\ref{Deltadefinition}),  between the
average energy of solar-produced neutrinos versus laboratory-produced
neutrinos.
Table~VI presents results of numerical calculations that
have been carried out by averaging the effect of vacuum oscillations
over the energy profile of the $^7$Be line that is computed using
the Bahcall-Pinsonneault solar model with helium diffusion (see Sec.\
\ref{sec:solarmodels}).
The quantity $\Delta(\nu_{e})$ that is given in Table~VI is
\begin{equation}
\Delta \left(\nu_e\right) \equiv
\frac{\int dq_{\rm obs} {\rm Spectrum} \left(q_{\rm
obs}\right)\left(1 - \sin^2 2\Theta_V \sin^2\phi \left(q_{\rm
obs}\right)\right)\left(q_{\rm obs} - q_{\rm lab}\right)}{\int d q_{\rm
obs} {\rm Spectrum}\left(q_{\rm obs}\right)\left(1 - \sin^2 2\Theta_V
\sin^2\phi\left(q_{\rm obs}\right)\right)}~,
\label{BigDeltae}
\end{equation}
which is the energy shift that would be measured for electron-type
neutrinos (e.g., in neutrino absorption experiments).  The corresponding
energy shift, $\Delta(\nu_{\mu})$, that would be measured with muon-type
neutrinos is
\begin{equation}
\Delta\left(\nu_\mu\right) \equiv \frac{\int dq_{\rm obs} {\rm
Spectrum}\left(q_{\rm obs}\right) \sin^2 \phi \left(q_{\rm
obs}\right)\left(q_{\rm obs} - q_{\rm lab}\right)}{\int d q_{\rm obs}
{\rm Spectrum}\left(q_{\rm obs}\right)\sin^2\phi\left(q_{\rm
obs}\right)}~.
\label{BigDeltamu}
\end{equation}
For neutrino-electron scattering experiments, the energy shift that
would be measured, $\langle\Delta\rangle_{\rm el. sc.}$, is
\begin{equation}
\left\langle\Delta\right\rangle_{\rm el.sc.} \equiv \frac{\sigma_e {\rm
Survival}\left(\nu_e\right)\Delta_e + \sigma_\mu\left(1 - {\rm
Survival}\left(\nu_e\right)\right)\Delta_\mu}{\sigma_e {\rm
Survival}\left(\nu_e\right) + \sigma_\mu\left(1 - {\rm
Survival}\left(\nu_e\right)\right)}~.
\label{BigDelaESc}
\end{equation}

I list in Table~VI representative numerical
results obtained for the two extreme values of the vacuum mixing
angle,
$\sin^2 2\Theta_V  = 1.00$ and $\sin^2 2\Theta_V  = 0.75$
[cf. Eq.\ (\ref{allowedsinsquared})];
intermediate choices of the vacuum mixing
angle yield intermediate effects that can be approximately interpolated
from Table~VI.
I do not list values when the
component in question carries less than 10\% of the flux,
since these cases are not relevant for currently feasible experiments.

The numerical calculations show that the energy shift, $\Delta$, for
both electron and muon neutrinos is within 10\% of the shift, 1.29 keV,
calculated in the absence of vacuum oscillations for all cases in which
the corresponding neutrino flux (electron or muon) is not less than 10
\% of the total flux.
A change of a given sign
in the energy shift for electron neutrinos, $\Delta(\nu_e)$,
implies a change of the opposite sign in the energy shift
for muon neutrinos, $\Delta(\nu_{\mu})$.
The calculated energy shift, $<\Delta>_{\rm el. sc.}$,
which should be used in making comparisons with
electron-neutrino scattering experiments,
is always within 3\% of the no-oscillation value of 1.29~keV.

\subsection{An MSW Solution}
\label{sec:mswsolution}

The MSW solution, matter-induced neutrino oscillations,
to the solar neutrino problem has been discussed by
many authors\cite{Wolfenstein78,Mikheyev86,Pal92}.
For matter-enhanced oscillations, the
probability of neutrino mixing within the sun depends upon neutrino
energy, but not in as delicate a fashion as for vacuum oscillations.
The dependence of the survival probability
on energy can be represented as a smooth function of energy over the extent
of the $^7$Be line profile\cite{Bahcall89,Wolfenstein78,Mikheyev86,Pal92}.
The dimensionless ratio that determines the amount by which the MSW
effect changes the line profile is the
ratio of the neutrino line width to the characteristic energy of the
line.  This ratio is very small,
${\rm FWHM}/2 q_{\rm peak} = 0.1\% $.

Large suppressions of electron-type neutrinos are achieved by the MSW
effect without fine-tuning with respect to the neutrino energy. In fact,
it does not seem possible to fine-tune MSW solutions to an accuracy that would
greatly amplify the small dimensionless ratio of FWHM to $q_{\rm peak}$
and thereby
affect significantly the line profile of a large fraction of
the $^7$Be solar neutrino flux.
In the resonance condition, the neutrino  energy, $q_{\rm res}$,
is inversely proportional to the electron density,$n_{\rm e}$,
at the resonance position\cite{Bahcall89,Wolfenstein78,Mikheyev86}. The
change in the electron density over the region of production of $^7$Be
neutrinos is\cite{Bahcall89}
$\Delta n_{\rm e} /n_{\rm e} \approx 0.25 $, while the
change in neutrino energy over the line profile is $\Delta q /q <
0.001$. If one tried to invent a situation in which only part of the
$^7$Be line profile went through resonance, the fine-tuning that applied
at one location would
quickly be destroyed as the electron density changed by a tiny amount
(and therefore changed the resonance condition)
within the region in which the neutrinos are produced.

\subsection{Some Other New Physics Processes}
\label{sec:otherprocesses}

Other solutions have been proposed for the solar neutrino problem that involve
new weak interaction physics.  These other solutions include
rotation of the neutrino magnetic moment \cite{Cisneros71}, matter-enhanced
magnetic moment transitions \cite{Lim88}, and neutrino decay
\cite{Bahcall72}.  The classical magnetic moment transition is
independent of energy and does not affect the shape of the line profile.
Matter-enhanced magnetic-moment transitions, like MSW transitions,
are not fine-tuned, vary smoothly with energy, and
depend upon a resonant electron density that varies from point to
point.
Therefore, the argument given above for the MSW effect also applies to
matter-enhanced magnetic-moment transitions.
Neutrino decay involves
characteristic energies that are very large compared to
the total width of the $^7$Be line and is also not fine-tuned.  Hence,
none of these processes would change significantly the shape of the
$^7$Be line profile.

\section{The $^7$Li Neutrino Absorption Cross Section}
\label{sec:Li7neutrino}

The calculated absorption cross section for the reaction
\begin{equation}
\nu_e + {\rm ^7Li} \to {\rm ^7Be} + e^- ,~~~~{\rm E_{th} = 861.9\
keV} ,
\label{Be7abs}
\end{equation}
where $\nu_e$ is produced by $^7$Be electron capture in the sun, depends
upon the assumed energy profile of the solar neutrinos
\cite{Domogatsky69,Bahcall78}. Neutrinos with energies below the energy
threshold, $E_{\rm th}$,
for the laboratory reaction, Eq.\ (\ref{Be7abs}), cannot be
absorbed. The energy threshold of 861.90 keV (cf. Eq.\ \ref{sumlab})
falls within the line profile
shown in Fig. II.
The precise location of the threshold
within the line profile determines the fraction of
emitted neutrinos that can be absorbed by $^7$Li.

The average absorption cross section for solar-produced $^7$Be neutrinos
incident on a laboratory detector of $^7$Li is
\begin{equation}
\langle{\rm Spectrum}_{\nu_e}(q_{\rm obs})\ \sigma_{\rm abs}(q_{\rm
obs})\rangle\simeq 19 \times 10^{-46}~{\rm cm}^2 ,
\label{crosssection}
\end{equation}
assuming neutrinos do not change flavor after their creation.
To the accuracy shown, the results are identical for $^7$Be line
profiles calculated using the standard
models\cite{Bahcall92} with and without helium diffusion (cf.  rows 1
and 2 of
Table~III).  As usual, Eq.\ (\ref{crosssection}) includes a correction to
take account of the fact that only 89.7\% of the $^7$Be neutrinos are
produced in ground-state to ground-state transitions.

The cross section given in Eq.\ (\ref{crosssection}) is almost a factor
of two larger than obtained previously \cite{Domogatsky69,Bahcall78},
which should make the contemplated radiochemical experiments somewhat
easier than originally considered \cite{Rowley78}.
The earlier treatments neglected the difference in electron binding
energies of solar and laboratory $^7$Be atoms as well as Doppler shifts of
the $^7$Be nuclei, and did not average over the temperature profile of
the sun.  In the present calculation, 88\% of the $^7$Be neutrinos are
above threshold for the reaction Eq.\ (\ref{Be7abs}).  For the earlier
calculation\cite{Bahcall78}, only about 50\% were above threshold.  The
difference between a cumulative probability of 50\% and a cumulative
probability of 88\% corresponds to an average energy shift of 0.85~keV
for the line profile shown in Fig.~2.

\section{Summary}
\label{sec:summary}

The temperature distribution of the solar core is expressed robustly in
the neutrino energy profile that results from $^7$Be electron capture.  The
characteristics of the line profile---shown in detail in Figures~2 and
3---are independent of uncertain details regarding solar
models and instead reflect the global thermal properties of the solar interior.
The robustness of the computed characteristics of the $^7$Be line
profile derives from the well-determined thermal structure of the
solar-model description of the interior
of the sun (cf. Sec.~\ref{sec:solarmodels}).  For the most precise
standard solar
models computed over the past decade, 1982--1992, and listed in
Table~III,
the central temperature has varied over a total range of $\pm 0.5$\%,
$T_c = (15.58 \pm 0.08) \times 10^6~{\rm K}$.
The $^7$Be and the $^8$B neutrino fluxes have varied by $\pm 6$\%.  The
relative number of electron captures that occur from bound orbits is
$0.221 \pm 0.006$ for all the solar models of Table~III.
For a heterogeneous set of ten recently-calculated
solar models, listed in Table~IV,
that were generally not required to
have the highest attainable precision, the central temperature varied by
$\pm 1$\%, $T_c = (15.55 \pm 0.15) \times 10^6~{\rm K}$.

The following paragraphs summarize the principal results obtained in
this paper.  In addition, Tables~II and III present numerical
representations of the
line profiles and Table~V provides a concise summary of the potentially
measurable characteristics that were calculated using different solar
models. The numerical values for the energy shift, $\Delta$, and the
low-energy and high-energy half-widths of the line profile, are
calculated in Sec.~\ref{sec:energyprofile}.

The shift in average neutrino energy,
$\Delta$, between $^7$Be neutrinos emitted in
the sun and $^7$Be neutrinos produced in the laboratory is $\Delta ({\rm
g.s.})= 1.29$~keV for ground-state captures and
the physical parameters of the most accurate available solar model
(which includes helium diffusion).  Calculations with other models yield
values of the
energy shift for ground-state decays
of 1.28~keV, 1.28~keV, and 1.23~keV (for a
1982 solar model with only 19 shells in the region in which $^7$Be
neutrinos are produced).  The shift for excited-state decays is
$\Delta ({\rm ex.s.}) = 1.24$~keV.  These calculated energy shifts
take account of
the fact that approximately 88\% of the captures involve electrons in
continuum orbits and only about 22\% involve electrons that are bound to
the decaying nucleus.  The atomic binding energies that are released
when $^7$Be nuclei capture electrons in the sun (or in the laboratory)
are evaluated in Sec.~\ref{sec:energies}.

The low-energy half-width, $W_-$, of the line profile is $(0.55 \pm
0.02)$~keV for the ground-state decay (0.24~keV for the excited-state
decay, see column 5 of Table V).
Here $W_-$ is the half-width of the energy profile below the peak in
the probability distribution.  This low-energy part of the line is
primarily determined by Doppler shifts caused by the thermal velocities
of $^7$Be ions that are moving away from the detector located on earth.
The low-energy side of the profile is approximately Gaussian in shape,
reflecting the Doppler origin of this part of the energy spectrum.

The high-energy half-width, $W_+$, of the line profile is
1.07~keV for the ground-state decay (0.73~keV for the excited-state
decay, see column~6 of Table~V) and is determined primarily by the
center-of-momentum kinetic energies of the electrons and the $^7$Be
nuclei that take part in electron capture reactions.  The high-energy
side of the line profile is approximately exponential in shape, with a
probability distribution that is proportional to
$\exp(- q_{\rm obs}/T_{\rm eff})$ where $q_{\rm obs}$
is the neutrino energy that is observed in the laboratory and $T_{\rm eff}
\approx 1.31~{\rm keV}\ (15.1
\times 10^6~{\rm K})$.  The exponential side of the probability
distribution results from an average (in the solar model) over the
different exponential distributions of center-of-momentum energies
(electrons and ions) that apply at each solar radius.

The principal characteristics of the line profile are derived by
approximate analytic calculations in Sec.~\ref{sec:approximate},
calculations that elucidate the physical origins of the various effects.
 The most remarkable result obtained in Sec.~\ref{sec:approximate} is
that the energy shift, $\Delta$, is essentially identical for the
ground-state and the excited-state captures (cf. Fig.~1).  The average
shift is shown in Sec.~\ref{sec:approximate} to be independent of the
typical emitted neutrino energy (384~keV or 862~keV) because the
 rest-mass energy of a $^7$Be nucleus is much larger than the solar
thermal energies (see discussion in Sec.~\ref{sec:energyshift}).

The energy shift is, to a good approximation, equal to
the average temperature of the
solar interior weighted by the fraction of $^7$Be neutrinos that are
produced at each temperature, i. e. $\int_\odot dTT d\phi({\rm
^7Be}, T)/\int_\odot dT d\phi({\rm
^7Be}, T)$, where $d\phi({\rm ^7Be},T)$ is the flux of $^7$Be
neutrinos produced at the local temperature $T$
[cf. Eq.\ (\ref{finale})]. The $^7$Be neutrinos are
produced in the inner few percent of the solar mass\cite{Bahcall89},
essentially all in the region $\left(0.04 \pm 0.03 \right)M_\odot$.
Therefore,  a measurement of the energy shift is a
measurement of the central temperature distribution of the sun.

The most striking aspect of the computed
energy profile is the asymmetry between the Gaussian
low-energy side
and the
exponential high-energy side
(cf. Figs.~2 and 3).
Doppler shifts caused by the thermal velocities of the $^7$Be nuclei produce a
symmetric, Gaussian contribution to the line broadening, which
determines the shape of the energy profile at energies below the peak.
The higher-energy part of the profile is determined by the
center-of-momentum kinetic energies. The exponential distribution
of kinetic energies produces an exponential
tail for large neutrino energies.
The positive-definite
character of the kinetic
energies is responsible for the asymmetry.

Electrostatic screening,
particle collisions, and gravitational redshifts all contribute to the
line broadening, but their effects are much smaller than
the effect caused by thermal broadening (Sec.~\ref{sec:othereffects}).

Vacuum neutrino oscillations can be fine-tuned to produce maximal mixing
of neutrino flavors near the peak of the $^7$Be neutrino line.  But, the
energy shift of the dominant neutrino survivor is always close to
the unmixed value of $\Delta = 1.29$~keV (see Table~VI).  The
invariance of the line shape results from the fact that
the oscillation phase changes
significantly only over an energy range that is much larger than the
line shape (see Sec.~\ref{sec:oscillations}).

The energy profile of the $^7$Be neutrino line should be taken
into account (using Tables~II and III) in precise calculations
of what is to be expected from
vacuum neutrino oscillations.  It has become standard in
calculations of the effects of vacuum oscillations
to take account of the variation of the
distance between the point of creation of the neutrinos
and the point of detection.  The variation in the point of
creation corresponds to a
phase-change of order $10^{-4}$, since the ratio of the
solar radius to the earth-sun distance is about $0.005$ and the
$^7$Be neutrinos are produced in a region of about $\pm 0.025
R_{\odot}$. On the other hand, the width of the $^7$Be line profile is
about $10^{-3}$ of the average $^7$Be neutrino energy.
Therefore, the change in phase due to the energy width of the neutrino
line is about an order of magnitude larger than the phase-change caused
by averaging over the region of production.

The MSW solutions that are consistent with
existing solar neutrino experiments
vary smoothly with energy and are not fine-tuned. The variation of the
resonant electron density within the sun
prevents fine-tuning of the solution
over a small energy range like the width of the
$^7$Be neutrino line.
For MSW solutions, the
fractional change in the electron-type neutrino survivability
over the energy profile of the $^7$Be line is
small. Other solutions of the solar neutrino problem that involve new
physics like rotation of the neutrino magnetic moment, matter-enhanced
magnetic-moment transitions, and neutrino decay are also not expected to
change significantly the shape of the $^7$Be line profile.

The energy shift and the shape of the $^7$Be neutrino energy profile
affect significantly the computed value of the absorption cross section
for $^7$Be solar neutrinos incident on a $^7$Li detector (see
Section~\ref{sec:Li7neutrino}).

\acknowledgments

This work was supported by NSF grant \#PHY92-45317.  I am grateful to
B.~Cabrera, F.~Calaprice, E.~Fiorini, M.~Giammarchi, A.~Gould,
M.~Kamionkowski, M. Lowry, P.~Krastev, R.~Kulsrud, A.~Loeb,
M.~N.~Rosenbluth, M. Schwarzschild, A.~Y.~Smirnov, E.~Witten, and T.
Ypsilantis for valuable discussions and suggestions.

% End references.

\begin{figure}
\caption[]{The $^7{\rm Be}$ Decay Scheme. The laboratory decay scheme for
$^7{\rm Be}$ is shown.  The neutrino energies emitted in the
ground-state to ground-state decay (branching ratio: 89.7\%) and
in the ground-state to excited-state decay
(branching ratio 10.3\%) are denoted,
respectively, by $q_{\rm Lab} ({\rm g.s.})$ and $q_{\rm Lab} ({\rm
ex.s.})$.  Details of the nuclear physics properties are summarized
in \protect\cite{Lederer78}.}
\label{fig1}
\end{figure}
\begin{figure}
\caption[]{The Energy Profile for the 862 keV line.  The probability for
the emission of a neutrino with energy $q_{\rm obs}$ in the laboratory
frame is shown as a function of $q_{\rm obs} - q_{\rm lab}$,
where the peak in the probability distribution is $q_{\rm peak} =
862.27$~keV, 0.43~keV larger than the laboratory decay energy, $q_{\rm
lab} = 861.84$~keV.
The probability distribution was computed by
averaging Eq.\ (\ref{rebasicrelation}),Eq.\ (\ref{boundenergyprofile}),
Eq.\ (\ref{fbound}),Eq.\ (\ref{allequations}),
Eq.\ (\ref{spectrumcombined}), and Eq.\ (\ref{averagespectrum}) over the
Bahcall-Pinsonneault standard solar model with helium diffusion.}
\label{fig2}
\end{figure}
\begin{figure}
\caption[]{The Energy Profile for the 384~keV line.  This figure was
computed in the same way as Fig.~2. The peak of this distribution lies
at $q_{\rm peak} = 384.47$~keV, 0.19~keV larger than the laboratory
decay energy, $q_{\rm lab} = 384.28$~keV.}
\label{fig3}
\end{figure}

\vfill\eject

\begin{table}
\tighten
\caption[]{The Energy Profile of the 862.0 keV $^7$Be solar neutrino line.
The neutrino energy is measured relative to the energy of the peak,
which occurs at $q_{\rm peak} = 862.27$~keV.  Here
$P(q_{\rm obs} - q_{\rm peak}) = {\rm Spectrum}_{\rm solar}(q_{\rm obs})$
is the
probability that a neutrino of energy $q_{\rm obs}$ will be emitted between
$q \pm
0.1$~keV.}
\begin{tabular}{dddddd}
\multicolumn{1}{c}{$q_{\rm obs}-q_{\rm peak}$}&$P(q_{\rm obs}-q_{\rm peak})$
&\multicolumn{1}{c}{$q_{\rm obs}-q_{\rm peak}$}&$P(q_{\rm obs}-q_{\rm peak})$
&\multicolumn{1}{c}{$q_{\rm obs}-q_{\rm peak}$}&$P(q_{\rm obs}-q_{\rm peak})$
\\
\hline
-2.100 & 0.00000 &  1.000 & 0.26096 &  4.100 & 0.02526\\
-2.000 & 0.00000 &  1.100 & 0.24204 &  4.200 & 0.02341\\
-1.900 & 0.00001 &  1.200 & 0.22456 &  4.300 & 0.02169\\
-1.800 & 0.00004 &  1.300 & 0.20838 &  4.400 & 0.02010\\
-1.700 & 0.00012 &  1.400 & 0.19339 &  4.500 & 0.01863\\
-1.600 & 0.00032 &  1.500 & 0.17946 &  4.600 & 0.01726\\
-1.500 & 0.00082 &  1.600 & 0.16655 &  4.700 & 0.01600\\
-1.400 & 0.00197 &  1.700 & 0.15455 &  4.800 & 0.01482\\
-1.300 & 0.00442 &  1.800 & 0.14342 &  4.900 & 0.01373\\
-1.200 & 0.00928 &  1.900 & 0.13308 &  5.000 & 0.01272\\
-1.100 & 0.01825 &  2.000 & 0.12347 &  5.100 & 0.01179\\
-1.000 & 0.03359 &  2.100 & 0.11455 &  5.200 & 0.01092\\
-0.900 & 0.05787 &  2.200 & 0.10627 &  5.300 & 0.01012\\
-0.800 & 0.09338 &  2.300 & 0.09858 &  5.400 & 0.00937\\
-0.700 & 0.14115 &  2.400 & 0.09145 &  5.500 & 0.00868\\
-0.600 & 0.20014 &  2.500 & 0.08482 &  5.600 & 0.00805\\
-0.500 & 0.26666 &  2.600 & 0.07867 &  5.700 & 0.00745\\
-0.400 & 0.33448 &  2.700 & 0.07295 &  5.800 & 0.00690\\
-0.300 & 0.39629 &  2.800 & 0.06765 &  5.900 & 0.00640\\
-0.200 & 0.44517 &  2.900 & 0.06274 &  6.000 & 0.00592\\
-0.100 & 0.47662 &  3.000 & 0.05817 &  6.100 & 0.00549\\
 0.000 & 0.48933 &  3.100 & 0.05394 &  6.200 & 0.00508\\
 0.100 & 0.48515 &  3.200 & 0.05001 &  6.300 & 0.00471\\
 0.200 & 0.46807 &  3.300 & 0.04636 &  6.400 & 0.00436\\
 0.300 & 0.44285 &  3.400 & 0.04298 &  6.500 & 0.00404\\
 0.400 & 0.41375 &  3.500 & 0.03984 &  6.600 & 0.00374\\
 0.500 & 0.38390 &  3.600 & 0.03693 &  6.700 & 0.00347\\
 0.600 & 0.35519 &  3.700 & 0.03423 &  6.800 & 0.00321\\
 0.700 & 0.32846 &  3.800 & 0.03173 &  6.900 & 0.00297\\
 0.800 & 0.30396 &  3.900 & 0.02941 &  7.000 & 0.00275\\
 0.900 & 0.28153 &  4.000 & 0.02726 & \\
\end{tabular}
\end{table}
\clearpage
\begin{table}
\tighten
\caption[]{The Energy Profile of the 384.5 keV $^7$Be solar neutrino
line. The neutrino energy is measured relative to the energy of the peak
which occurs at $q_{\rm peak} = 384.47$~keV.  Here
$P(q_{\rm obs} - q_{\rm peak}) = {\rm Spectrum}_{\rm solar}(q_{\rm obs})$ is
the
probability that a neutrino of energy $q_{\rm obs}$ will be
emitted between $q \pm
0.1$~keV.}
\begin{tabular}{dddddd}
\multicolumn{1}{c}{$q_{\rm obs}-q_{\rm peak}$}&$P(q_{\rm obs}-q_{\rm peak})$
&\multicolumn{1}{c}{$q_{\rm obs}-q_{\rm peak}$}&$P(q_{\rm obs}-q_{\rm peak})$
&\multicolumn{1}{c}{$q_{\rm obs}-q_{\rm peak}$}&$P(q_{\rm obs}-q_{\rm peak})$
\\
\hline
-1.000 & 0.00000 &  1.800 & 0.16166 &  4.600 & 0.01956\\
-0.900 & 0.00000 &  1.900 & 0.15004 &  4.700 & 0.01812\\
-0.800 & 0.00002 &  2.000 & 0.13924 &  4.800 & 0.01679\\
-0.700 & 0.00022 &  2.100 & 0.12921 &  4.900 & 0.01556\\
-0.600 & 0.00187 &  2.200 & 0.11989 &  5.000 & 0.01442\\
-0.500 & 0.01152 &  2.300 & 0.11124 &  5.100 & 0.01336\\
-0.400 & 0.05065 &  2.400 & 0.10320 &  5.200 & 0.01238\\
-0.300 & 0.15855 &  2.500 & 0.09574 &  5.300 & 0.01147\\
-0.200 & 0.35442 &  2.600 & 0.08881 &  5.400 & 0.01063\\
-0.100 & 0.57417 &  2.700 & 0.08238 &  5.500 & 0.00985\\
 0.000 & 0.69841 &  2.800 & 0.07641 &  5.600 & 0.00913\\
 0.100 & 0.68431 &  2.900 & 0.07087 &  5.700 & 0.00846\\
 0.200 & 0.60023 &  3.000 & 0.06573 &  5.800 & 0.00783\\
 0.300 & 0.52127 &  3.100 & 0.06095 &  5.900 & 0.00726\\
 0.400 & 0.46715 &  3.200 & 0.05652 &  6.000 & 0.00673\\
 0.500 & 0.42841 &  3.300 & 0.05241 &  6.100 & 0.00623\\
 0.600 & 0.39618 &  3.400 & 0.04859 &  6.200 & 0.00577\\
 0.700 & 0.36720 &  3.500 & 0.04505 &  6.300 & 0.00535\\
 0.800 & 0.34060 &  3.600 & 0.04177 &  6.400 & 0.00495\\
 0.900 & 0.31603 &  3.700 & 0.03872 &  6.500 & 0.00459\\
 1.000 & 0.29330 &  3.800 & 0.03590 &  6.600 & 0.00425\\
 1.100 & 0.27224 &  3.900 & 0.03328 &  6.700 & 0.00394\\
 1.200 & 0.25271 &  4.000 & 0.03085 &  6.800 & 0.00365\\
 1.300 & 0.23460 &  4.100 & 0.02859 &  6.900 & 0.00338\\
 1.400 & 0.21778 &  4.200 & 0.02650 &  7.000 & 0.00313\\
 1.500 & 0.20216 &  4.300 & 0.02457 &  7.100 & 0.00290\\
 1.600 & 0.18765 &  4.400 & 0.02277 &  7.200 & 0.00269\\
 1.700 & 0.17418 &  4.500 & 0.02110 & \\
\end{tabular}
\end{table}
\clearpage
\begin{table}
\tightenlines
\squeezetable
\caption[]{Some characteristics of four standard solar models.}
\begin{tabular}{lcccccc}
\multicolumn{1}{c}{Model}&Shells&$T_{\rm central}$&$\langle T\rangle$&Bound
Fraction&$\phi(^7{\rm Be})$&$\phi(^8{\rm B})$\\
&$(T > 6 \times 10^6~{\rm K})$&$(10^6~{\rm K})$&(Weight:\ $^7$Be)&$^7$Be
decays)&$(10^9~{\rm cm^{-2} s^{-1}})$&$(10^6~{\rm cm^{-2} s^{-1}})$\\
\hline
\noalign{\smallskip}
Bahcall-Pinsonneault 1992&127&15.67&14.13&0.215&4.9&5.7\\
\multicolumn{1}{c}{(Helim diffusion)}\\
\noalign{\medskip}
Bahcall-Pinsonneault 1992&127&15.57&14.06&0.217&4.6&5.1\\
\multicolumn{1}{c}{(No diffusion)}\\
\noalign{\medskip}
Bahcall-Ulrich 1988&87&15.64&14.07&0.219&4.7&5.8\\
\multicolumn{1}{c}{(No diffusion)}\\
\noalign{\medskip}
Bahcall et al. 1982&19&15.50&13.6&0.226&4.3&5.6\\
\multicolumn{1}{c}{(No diffusion)}\\
\end{tabular}
\end{table}
\begin{table}
\tightenlines
\caption[]{Central temperatures of 12 recently-computed solar
models (no diffusion).}
\begin{tabular}{lc}
\multicolumn{1}{c}{Model}&$T_c$ \\
&$(10^6~{\rm K})$ \\
\hline
\noalign{\smallskip}
Castellani, Degl'Innocenti, and Fiorentini (1993)&15.72\\
Berthomieu et al. (1993)&15.55 \\
Turck-Chi\`eze and Lopes (1993)&15.43 \\
Ahrens, Stix, and Thorn (1992)&15.65 \\
Bahcall and Pinsonneault (1992)&15.57\\
Christensen-Dalsgaard (1992)&15.68 \\
Guenther et al. (1992)&15.53 \\
Guzik and Cox (1991)&15.40 \\
Proffitt and Michaud (1991)&15.71 \\
Sackmann et al. (1990)&15.43 \\
Bahcall and Ulrich (1988)&15.64\\
Lebreton and D\"appen (1988)&15.54 \\
\end{tabular}
\end{table}
\begin{table}
\tightenlines\squeezetable
\caption[]{Characteristics of the energy profile.}
\begin{tabular}{lcccccccc}
\multicolumn{1}{c}{Model}&$\Delta$&$\langle q - q_{\rm peak}\rangle$
&FWHM&$W_-$&$W_+$&$\langle(q - q_{\rm peak})^2\rangle$&
$\langle(q - q_{\rm peak})^3\rangle$&$\delta q$ \\
&(keV)&(keV)&(keV)&(keV)&(keV)&(keV)&(keV)&(keV) \\
\hline
\noalign{\bigskip\smallskip}
\multicolumn{9}{c}{\underline{\ \ \ \ Ground-state decay\ \ \ \ }}\\
\noalign{\medskip\smallskip}
Bahcall-Pinsonneault 1992&1.29&0.856&1.63&0.56&1.07&2.59&10.0&0.23 \\
\ \ (Helium diffusion) \\
\noalign{\medskip}
Bahcall-Pinsonneault 1992&1.28&0.858&1.62&0.55&1.07&2.58&9.9&0.22 \\
\ \ (No diffusion) \\
\noalign{\medskip}
Bahcall-Ulrich 1988&1.28&0.857&1.62&0.55&1.07&2.58&9.9&0.22 \\
\ \ (No diffusion) \\
\noalign{\medskip}
Bahcall et al. 1982&1.23&0.821&1.56&0.53&1.03&2.39&8.8&0.21 \\
\ \ (No diffusion) \\
\noalign{\bigskip}
\multicolumn{9}{c}{\underline{\ \ \ \ Excited-state decay\ \ \ \ }}\\
\noalign{\medskip\smallskip}
Bahcall-Pinsonneault 1992&1.24&1.048&0.97&0.24&0.73&2.86&11.3&0.04 \\
\ \ (Helium diffusion) \\
\noalign{\medskip}
Bahcall-Pinsonneault 1992&1.23&1.040&0.96&0.24&0.72&2.82&11.1&0.04 \\
\ \ (No diffusion) \\
\noalign{\medskip}
Bahcall-Ulrich 1988&1.23&1.039&0.95&0.24&0.71&2.82&11.1&0.04 \\
\ \ (No diffusion) \\
Bahcall et al. 1982&1.18&1.002&0.91&0.22&0.69&2.62&9.9&0.03 \\
\ \ ((No diffusion) \\
\end{tabular}
\end{table}
\begin{table}
\tighten
\caption[]{The effect of vacuum neutrino oscillations on the energy
shift, $\Delta (\nu_e)$ [see Eq.\ (\ref{ReDeltadefinition})], for
electron neutrinos and on the energy shift , $\Delta (\nu_\mu)$,
for muon neutrinos is given
as a function of vacuum mixing angles, $\Theta_V$, and the phase-angle,
$\phi$, see Eq.\ (\ref{phivacuumvalue}), evaluated at the peak of the
energy spectrum at which neutrinos are created, $q_{\rm peak}=
862.27$~keV. The last column gives the energy shift, $\Delta$(e.sc.),
that would be measured in an electron-scattering experiment.
The survival probability, averaged over the neutrino
profile, for an electron neutrino to remain an electron neutrino is
given in column 3.}
\begin{tabular}{dddddd}
$\sin^2 2\Theta_V$&$2\phi(q_{\rm peak})/\pi$&Survival $(\nu_e)$&$\Delta
(\nu_e)$&$\Delta(\nu_\mu)$&\multicolumn{1}{c}{$\Delta$(e.sc.)}\\
&&&(keV)&(keV)&(keV)\\
\hline
0.0&0.0&1.000&1.286&$\cdots$&1.29\\
1.00&3.5&0.495&1.26&1.31&1.27\\
1.00&4.0&0.9999&1.29&$\cdots$&1.29\\
1.00&4.5&0.507&1.32&1.26&1.31\\
1.0&4.6&0.352&1.33&1.26&1.31\\
1.0&4.7&0.212&1.35&1.27&1.31\\
1.0&4.8&0.100&1.39&1.27&1.31\\
1.0&4.9&0.027&$\cdots$&1.28&1.30\\
1.0&5.0&0.0002&$\cdots$&1.29&1.29\\
1.0&5.1&0.022&$\cdots$&1.29&1.27\\
1.0&5.2&0.091&$\cdots$&1.30&1.26\\
1.0&5.3&0.200&1.22&1.30&1.26\\
1.0&5.4&0.338&1.24&1.31&1.26\\
1.0&5.5&0.492&1.25&1.32&1.26\\
1.00&6.0&0.9997&1.29&$\cdots$&1.29\\
1.00&7.0&0.0004&$\cdots$&1.29&1.29\\
1.0&7.2&0.089&$\cdots$&1.300&1.25\\
1.0&7.4&0.335&1.22&1.32&1.25\\
1.0&7.6&0.643&1.25&1.36&1.26\\
1.0&7.8&0.897&1.27&1.45&1.27\\
1.0&8.0&0.9995&1.28&$\cdots$&1.28\\
0.75&3.5&0.621&1.27&1.31&1.28\\
0.75&4.0&0.9999&1.29&$\cdots$&1.29\\
0.75&4.5&0.630&1.30&1.26&1.30\\
0.75&5.0&0.250&1.29&1.29&1.29\\
0.75&6.0&0.9998&1.29&$\cdots$&1.29\\
0.75&7.0&0.250&1.29&1.28&1.29\\
0.75&8.0&0.9996&1.29&$\cdots$&1.29\\
\end{tabular}
\end{table}
\end{document}